\documentclass[12pt]{article}

\usepackage[utf8]{inputenc} 
\usepackage[T1]{fontenc} 
\usepackage[osf]{newpxtext} 
\usepackage[varg,bigdelims,cmintegrals,cmbraces]{newpxmath} 
\usepackage{textcomp} 
\usepackage[scr=rsfso]{mathalfa} 
\usepackage{bm} 

\usepackage{geometry} 
\usepackage{afterpage}

\usepackage{setspace} 
\usepackage{graphicx} 
\usepackage{tabularx, ragged2e} 
\usepackage{multirow} 
\usepackage{adjustbox} 
\usepackage{svg}
\newcolumntype{Y}{>{\centering\arraybackslash}X} 
\newcolumntype{b}{X} 
\newcolumntype{s}{>{\hsize=.5\hsize}X} 

\usepackage{amsmath,latexsym} 

\usepackage[backend=bibtex,style=authoryear,maxbibnames=99,maxcitenames=2]{biblatex} 
\DeclareNameAlias{sortname}{last-first} 

\usepackage{hyperref} 
\hypersetup{
    colorlinks,
    linkcolor={red!50!black},
    citecolor={blue!50!black},
    urlcolor={blue!80!black},
    pdftitle={},
    pdfauthor={},
    pdfsubject={},
    pdfkeywords={},
    bookmarksnumbered=true,
    bookmarksopen=true,
    bookmarksopenlevel=1,
    pdfstartview={XYZ null null 0.75},
    pdfpagemode=UseOutlines,
    pdfpagelayout=OnePageRight
}

\usepackage{floatrow}
\usepackage{subfig}
\usepackage{caption}
\usepackage{xcolor} 
\usepackage{dsfont} 
\usepackage{pdflscape} 
\usepackage{lscape} 
\usepackage{float} 
\usepackage{multicol} 
\usepackage{booktabs} 
\usepackage[bottom]{footmisc} 

\definecolor{darkblue}{rgb}{0,0,0.5}
\definecolor{mauve}{rgb}{0.88,0.69,1}

\usepackage[affil-sl]{authblk}
\makeatletter
\renewcommand\AB@authnote[1]{{\normalfont\textsuperscript{#1}}}
\renewcommand\AB@affilnote[1]{{\normalfont\textsuperscript{#1}}}
\makeatother

\usepackage[normalem]{ulem}
\providecolor{added}{rgb}{0,0,1}
\providecolor{deleted}{rgb}{1,0,0}

\title{The Mortgage Cash-Flow Channel: How Rising Interest Rates Impact Household Consumption\thanks{We thank Sigal Ribon, Boris Hofmann, Alon Raviv, Nittai Bergman and the participants in The Bank of Israel Research Department Seminar, the Tel Aviv University Applied Economics Workshop, the Annual International Journal of Central Banking Research Conference, the Annual Conference of the Israeli Economics Association, and the Ben-Gurion University Economics Department Seminar for very helpful comments and suggestions. The views expressed here are the authors' and do not necessarily reflect those of the Bank of Israel.}}
\author[1]{Itamar Caspi\thanks{Email: \url{itamar.caspi@boi.org.il}.}}
\author[2]{Nadav Eshel\thanks{Email: \url{nadav.eshel@boi.org.il}.}}
\author[1]{Nimrod Segev\thanks{Corresponding author. Email: \url{nimrod.segev@boi.org.il}.}}
\affil[1]{{\normalsize{Research Department, Bank of Israel}}}
\affil[2]{{\normalsize{Governor's Office, Bank of Israel}}}

\begin{document}
\singlespacing
\vspace{-2cm}
\maketitle
\vspace{-1cm}
\begin{abstract}
\noindent This study investigates the impact of increased debt servicing costs on household consumption resulting from monetary policy tightening. It utilizes observational panel microdata on all mortgage holders in Israel and leverages quasi-exogenous variation in exposure to adjustable-rate mortgages (ARMs) due to a regulatory shift. Our analysis indicates that when monetary policy became more restrictive, consumers with a higher ratio of ARMs experienced a more marked reduction in their consumption patterns.This effect is predominantly observed in mid- to lower-income households and those with a higher ratio of mortgage payments to total spending. These findings highlight the substantial role of the mortgage cash-flow channel in monetary policy transmission, emphasizing its implications for economic stability and inequality.
\newline

\vspace{0.1cm}

\noindent
\textit{Keywords: }Adjustable-rate mortgages, monetary policy, cash-flow channel, household consumption, heterogeneity, Israel. \newline
\noindent \textit{JEL Classification:} E30, E44, E58, G21. 

\vspace{20cm}

\setcounter{page}{0}
\thispagestyle{empty}
\pagebreak

\end{abstract}

\begin{quote}
``\textit{[T]he structure of mortgage contracts may matter for consumption behavior. In countries...where most mortgages have adjustable rates, changes in short-term interest rates have an almost immediate effect on household cash flows... In an economy where most mortgages carry fixed rates...that channel of effect may be more muted... [T]hese issues seem worthy of further study...}''

\raggedleft
--- \textcite{bernanke2007financial}
\end{quote}

\onehalfspacing
\section{Introduction}
The surprising global resurgence of inflation during 2021-2023 has lead many central banks worldwide to enact an aggressive tightening cycle not seen in decades. One of the key challenges for monetary policymakers is understanding the impact of policy rate changes on consumer spending, particularly in the context of adjustable-rate mortgages (ARMs). ARMs differ from fixed-rate mortgages in that the interest rate on the loan is periodically adjusted based on a benchmark or index, such as the central bank's policy rate. As a result, changes in monetary policy can directly influence the monthly mortgage payments of ARM holders, potentially affecting their disposable income and spending.

A country's mortgage market characteristics, such as the proportion of mortgages eligible for refinancing and the overall debt-servicing costs relative to household income, can significantly influence the transmission of monetary policy to household consumption and to the rest of the economy \autocite{di2017interest, agarwal2022mortgage, eichenbaum2022state}. More specifically, the prevalence of adjustable-rate mortgages (ARMs) in the housing market can affect the impact of interest rate changes on consumer spending. When a central bank raises interest rates, mortgagors with ARMs experience an immediate increase in their monthly mortgage payments, reducing their disposable income and potentially leading to a contraction in consumer spending. Conversely, in countries where fixed-rate mortgages dominate, the impact of interest rate changes on household budgets is more gradual, as only new or refinanced mortgages are affected. Thus, the share of ARMs in a country's mortgage market can affect the speed and magnitude of the consumption response to monetary policy actions, with important implications for the effectiveness of central bank measures in managing inflation and economic stability \autocite{imf2024world}.


Understanding the relationship between central bank rates and household expenditures is important for effective economic policy, especially in economies where mortgages play a significant role in personal finance. However, analyzing this impact is challenging due to the complex interplay between policy rates, mortgage payments, and consumer spending behavior. Our study uncovers the causal relationship between the exposure of households to interest changes via their exposure to variable rate mortgages and consumption. This relationship explains the observed simultaneous increase in mortgage payments and decrease in credit card spending. By examining these effects, we provide insight into an important channel through which monetary policy operates - the Mortgage Cash-Flow Channel.



Establishing the causal effect of mortgage cash flow on household consumption rests on two key elements: a proprietary dataset covering the entirety of Israeli mortgages and an empirical strategy that utilizes variations in exposure to interest rate changes. This approach is further strengthened by a natural experiment stemming from a regulatory change. We leverage the characteristics of Israeli mortgages to analyze the impact of policy rate changes on consumption expenditures. Moreover, our dataset allows us to examine how households adjust their spending in response to the negative cash-flow shock resulting from the Bank of Israel's interest rate hikes during 2022-2023. Our primary focus is on the direct effects of these rate increases on mortgage payments and these households' subsequent deferred debit spending patterns. Given that deferred debit are the primary payment method in Israel, 
this extensive dataset facilitates a comprehensive analysis of a significant part of mortgage households' consumption patterns during monetary contraction.

The empirical approach of this study employs a difference-in-differences (DiD) methodology to address potential endogeneity. The study uses a multi-stage analytical framework, initially categorizing borrowers based on the relative size of their ARM mortgage share, which is directly linked to the Bank of Israel's policy rate.

Additionally, we leverage a quasi-natural experimental setting, capitalizing on a regulatory change by the Bank of Israel in January 2021 that increased the ARM mortgage share cap from 33\% to 67\%. By focusing on the upper terciles of borrowers affected by this regulatory shift, we can compare borrower cohorts that are nearly identical, primarily differing in the magnitude of their ARM mortgage exposure. We argue that this comparison enables a complementing assessment of the interest rate change's impact on disposable income and spending patterns.


We find that a one percentage point higher ARM ratio causes the reduction in household spending by 4.6 basis points during a monetary tightening cycle in which the monetary policy rate increased from 0.1\% to 4.75\% within 15 months. Based on our estimated effect and data on the average increase in mortgage payments during the hiking period, we estimate a marginal propensity to consume (MPC) of around 0.4 for the average mortgage household. To put this into perspective, consider a household with an average ARM ratio of 32\%. The decline in overall spending for this household was 1.5\% greater compared to a household with no ARM exposure. Aggregating these effects to the national level underscores the macroeconomic significance of the mortgage cash flow channel. Finally, we estimate that the overall impact of the mortgage-cash flow channel on private consumption in Israel was a reduction of around 20 basis points for every 100 basis points increase in the policy rate.



Our analysis shows that among mortgage holders, the impact of higher exposure to adjustable-rate mortgages (ARMs) on household consumption is most significant for mid to lower income households and those with a higher proportion of mortgage spending compared to total spending before the period of monetary tightening. This finding suggests that changes in mortgage rates can disproportionately affect certain socioeconomic groups, potentially contributing to monetary policy redistribution effects \autocite{auclert2019monetary}.


Compared to previous studies that focused on specific samples or aggregated data, to the best of our knowledge, this paper is the first to use the universe of mortgages in a country to directly estimate the impact of the mortgage cash-flow channel. This paper's extensive dataset and identification strategies offer an in-depth understanding of the relationship between mortgage structures and interest rates. The findings stress the important role of the mortgage market structure in transmitting monetary policy and emphasize the need for policymakers to consider household debt dynamics when making decisions. Overall, this study contributes to a better grasp of the interactions between the housing market, household debt, and monetary policy, highlighting the significance of these factors in developing effective economic policies.

\vspace{0.2cm} \noindent {\textit{Related Literature.}}\quad 
This paper is related to a number of strands of the literature. First, we relate to the vast literature on the broad monetary transmission channels. Monetary policy exerts its influence on the broader economy through various channels. Traditional economic models generally view the impact of policy on the economy as being driven by changes in interest rates, a concept known as the "interest rate channel of monetary policy." These models typically operate under the assumption that prices and wages are relatively inflexible in the short term. Therefore, an increase or decrease in nominal interest rates leads to correspondingly higher or lower real interest rates. This change affects the cost of borrowing due to the intertemporal substitution effect, influencing both investment and consumer spending. However, the actual response of the economy to interest rate adjustments by the central bank is often more significant than these models suggest, indicating that there are other mechanisms at play beyond the direct influence of interest rates on the economy.\footnote{Research has identified several channels through which monetary policy affects the economy. These include: the Credit Channel \parencite{bernanke1995inside}, Exchange Rate Channel \parencite{eichenbaum1995some}, Expectations Channel \parencite{svensson1997inflation}, Bank Lending Channel \parencite{kashyap1994monetary}, Risk-Taking Channel \parencite{borio2012capital}, and Asset Price Channel \parencite{bernanke2005explains}. \textcite{choi2024revisiting} provides recent international evidence on the relative importance of these channels.}

 More specifically, this paper relates to the studies that examine a monetary mortgage Cash Flow Channel, especially in the context of ARMs and household consumption. This channel examines the impact of monetary policy changes on the cash flow of borrowers and consumers. Theoretically, with a high share of ARM mortgages, changes in interest rates can directly influence borrowers' cash flow by altering their debt servicing costs, thereby affecting their disposable income and their ability to spend and invest \autocite{garriga2017mortgages}.\footnote{More broadly, we can think of a debt service ratio (DSR) channel of monetary policy that relates to the association between debt (variable rate payments) of firms or households relative to income and macroeconomic outcomes (e.g., \textcite{hofmann2017there, ippolito2018transmission, cloyne2020monetary})}

Recent empirical studies provide support for the crucial role of the mortgage cash flow channel in monetary policy transmission and its influence on economic activities. Using aggregated country-level data, \textcite{calza2013housing} show that the presence of fixed-rate mortgages can significantly dampen the impact of a monetary policy shock on both consumption and residential investment. Using survey data on household-level expenditures, \textcite{cooper2021mortgage} compare the impact of interest rate changes in the USA, where mortgages are primarily fixed-rate, to Spain where they are primarily ARMs. The two papers that are most closely related to ours are \textcite{agarwal2022mortgage} and \textcite{di2017interest}. \textcite{agarwal2022mortgage} use a sample of credit card holders from a single Chinese bank to study the consumption response to mortgage payment reductions. Comparing the consumption response of mortgage versus non-mortgage consumers, the authors find a significant increase in credit card spending following a one-time 230 basis points drop in interest rates. \textcite{di2017interest} use the variation in the timing of expected interest rate resets for a sample of five-year ARMs. They find that reduction in mortgage payments, even when the reset date is entirely expected, induces a significant consumption response, with a specific focus on car purchases.

We differ from these studies in a number of ways. First, our paper looks at the impact of a largely unexpected interest rate increase on spending in a period of significant and continuous monetary contraction, while previous studies have focused on either reduction or expected mortgage payment changes. Second, we utilize cross-sectional variation between mortgages in their ARM exposure as well as a unique regulatory change to quantify the causal impact of changes in mortgage payments that are directly linked to changes in policy interest rates. Finally, we are the first to use household-level data for the entire population of mortgage holders in a country.

More broadly, our paper is related to the household finance literature which examines how household indebtedness, interest rates, access to home equity, and mortgage refinancing decisions impact household behavior \autocite{mian2013household, keys2016failure, agarwal2017access, agarwal2021interest} and the theoretical literature on the aggregate implications of the housing market \autocite{rubio2011fixed, guerrieri2017collateral, slacalek2020household, eichenbaum2022state}.

Finally, we contribute to the literature emphasizing the role of household heterogeneity in economic outcomes \autocite{krueger2016macroeconomics}. In particular, there is a growing recognition that various monetary policy channels, including the less examined Cash Flow Channel, can have differing impacts across diverse household groups. \textcite{kaplan2014wealthy} introduced the concept of a 'hand-to-mouth' population within macroeconomic models. They demonstrated that the consumption patterns of liquidity-constrained households significantly influence aggregate demand responses to fiscal policy changes. Similarly, \textcite{auclert2019monetary} highlighted the necessity of viewing monetary policy through a distributional lens, acknowledging the substantial macroeconomic effects of income and wealth disparities. Furthermore, the literature delves into income distribution as well as racial and regional variations in these effects, stressing the significance of local economic structures and inequality in the mechanism of monetary policy transmission \autocite{holm2021transmission, cumming2019, gerardi2023mortgage, tzamourani2021interest}. 

\vspace{0.2cm} \noindent {\textit{Outline.}}\: 
The remainder of the paper is organized as follows: Section~\ref{Section: inst back} provides an overview of monetary policy and the mortgage market in Israel. Section~\ref{Section: Data} describes the unique dataset from Israel and presents descriptive statistics. Section~\ref{Section: empirical} details the methodological approach. Section~\ref{Section: Results} discusses the empirical results and includes initial robustness checks. Section~\ref{Section: robustness} describes a battery of additional robustness tests. Finally, the paper concludes with Section~\ref{Section: Conclusions}, which summarizes the findings and implications of the study.

\section{Macroeconomic and Institutional Background}
\label{Section: inst back}

\subsection{Monetary Policy}
The Bank of Israel has three primary objectives as outlined in the 2010 Bank of Israel Law. Firstly, the Bank aims to maintain price stability, which is defined by the government as keeping the annual Consumer Price Index (CPI) inflation rate within a range of 1-3\%. Secondly, the Bank seeks to support government policies that contribute to economic growth, employment, and reducing inequality, as long as such support does not undermine its price stability objectives. Finally, the legislation mandates the Bank to preserve the stability of the financial system and ensure orderly market functioning (Bank of Israel Law, 2010). Therefore, while price stability is the Bank's primary objective, the law also acknowledges a secondary role for counter-cyclical policies and financial oversight.

To achieve these objectives, the Bank of Israel employs various policy instruments, with the primary tool being the short-term interest rate, set by a Monetary Committee led by the Governor. This rate, known as the Bank of Israel interest rate, determines the interest the Bank pays to commercial banks on their deposits. The Bank operates within a "flexible inflation target" where it sets this interest rate at a level that either maintains inflation within the target range or is expected to bring inflation back to the target range within a period not exceeding two years. The Bank operates independently in setting the short-term interest rate and in utilizing monetary instruments to achieve its goals.

\begin{figure}[!htb]
\centering
\includegraphics[width=0.8\textwidth]{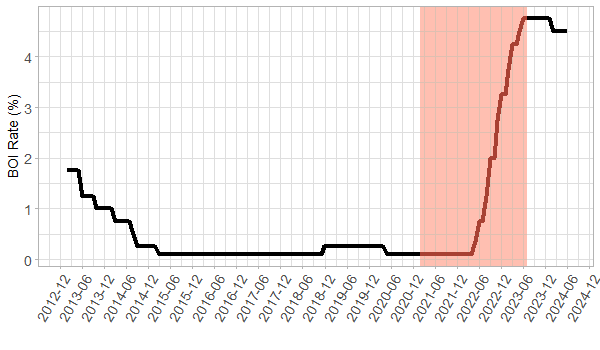}
\caption{Bank of Israel Monetary Policy Rate, 2012-2024)}
\label{Fig:BOI_Rate}
\floatfoot*{\textit{Notes}: This figure presents the Bank of Israel's policy rate. The shaded area denotes the sample period we use in our baseline estimation (see Section~\ref{Section: Data} for further details.)}
\end{figure}

Between April 2022 and July 2023, Israel's monetary policy underwent significant shifts in response to changing inflation dynamics, reflecting domestic and global economic factors. Throughout this period, Israel experienced an elevated inflation environment, with year-over-year inflation rates consistently surpassing the upper bound of the central bank's target range. The Bank of Israel attributed this rise in inflation primarily to increased domestic demand, alongside global economic trends. In response, the Bank of Israel embarked on a path of monetary tightening, primarily through a series of interest rate hikes. From mid-2022 to early 2023, the central bank increased the interest rate from 0.10\% to 3.75\%, continuing the tightening process initiated in the post-COVID-19 period (see Figure~\ref{Fig:BOI_Rate}.) This policy stance was further intensified in the first half of 2023, with the interest rate reaching 4.75\% by July 2023, following which the rate was kept there until the first rate cut in January 2024.\footnote{\textcite{boi_2022_second_half, boi_2023_first_half}.}

\subsection{The Israeli Mortgage Market}
\label{Section: mort Isr}
In Israel, banks are practically the sole originators of all mortgage products.\footnote{Although a few non-bank lending platforms provide mortgages, their combined market share as of 2023 accounts for less than two percent of both new and outstanding mortgages.} A typical Israeli mortgage is characterized by various "tracks," each having its specific interest rate structure. There are primarily three common interest rate options in the Israeli mortgage market: Long Fixed rate, where the mortgage interest rate is fully fixed; Medium-Fixed, where the base rate adjusts in accordance with government bonds every five years;\footnote{A minor portion adjusts biannually; however, their contribution to the mortgage balance in our sample is negligible.} and Variable, which includes mortgages directly tied to the Bank of Israel's interest rate.\footnote{In Israel the variable rate tracks are known as the ``Prime rate" track. This is not to be confused with the creditworthiness of the borrower but rather to the interest rate adjusting process.} In what follows, we will refer to the latter track as ARM and use it as our measure of exposure of mortgage borrowers to changes in the interest rate. Additionally, Fixed and Medium-Fixed tracks in Israel may also be linked to the price index.


As of September 2023, the total balance of mortgages in Israel is approximately NIS 540 billion (roughly 100\% of GDP). The distribution among the different tracks is quite telling: Fixed interest rates constitute 39 percent of the total credit, Medium-Fixed rates account for 21 percent, ARMs account for 39 percent, and the remaining 1 percent is in other less common interest tracks. Additionally, 36\% of the total credit is linked to the price index, with the bulk of those in the Medium-Fixed track.

\begin{figure}[!htb]
\centering
\caption{Mortgages Interest Rate Structure}
\includegraphics[width=0.8\textwidth]{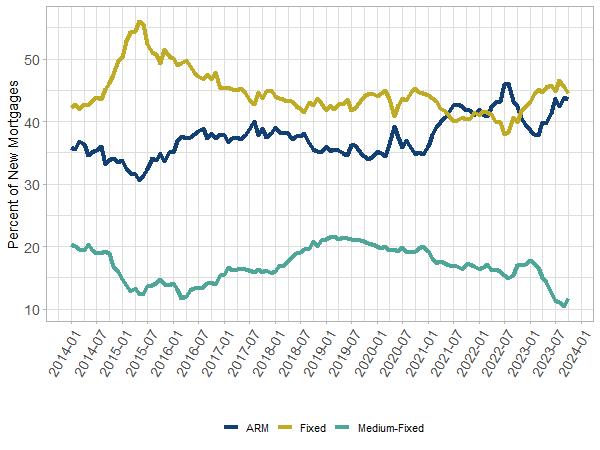}
\label{Fig mort structure}
\floatfoot*{\textit{Notes}: This figure presents the distribution of new mortgages in Israel categorized by their interest rate types. 
}
\end{figure}

\begin{figure}[!htb]
\centering
\caption{ARM Ratio Density}
\includegraphics[width=0.8\textwidth]{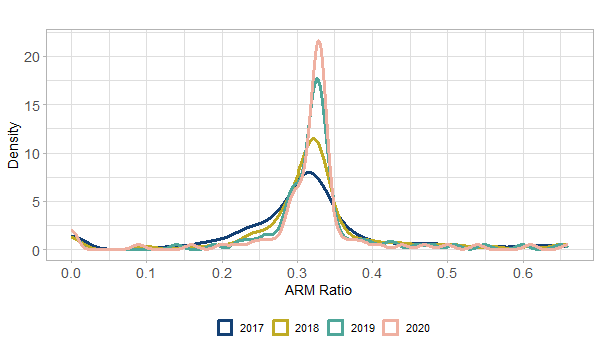}
\label{Fig mort density}
\floatfoot*{\textit{Notes}: This figure presents the density of the ARM track share from each mortgage \emph{current} balance in March 2022, split by mortgage origination year.  
}
\end{figure}

Mortgage borrowers in Israel have the option to tailor their mortgage structures by utilizing a mix of different financial products. Regulations stipulate that a minimum of one-third of the mortgage must be comprised of a fixed rate. Until January 2021, the proportion that could be allocated to the ARM track was capped at one-third. Figure \ref{Fig mort structure} depicts the share of each mortgage type within the new mortgage balances over the previous decade, showing notable variances in mortgage structures over this period.\footnote{It is essential to recognize that the regulation is only applicable at the inception of the mortgage. Given that various tracks within a single mortgage may have different maturity dates, and borrowers have the discretion to repay specific tracks preferentially, the proportion of the ARM track within the outstanding balance may surpass one-third, while the fixed rate's share may fall below this threshold.} Furthermore, there is substantial diversity in mortgage configurations across various segments. For example, Figure \ref{Fig mort density} describes the distribution of the ARM share of all mortgages in March 2022, segmented by four different origination year of the mortgages. This information underscores pronounced disparities in the direct exposure to the Bank of Israel rate via the ARM track among cohorts of mortgage borrowers with loans initiated in proximate time-frames.\footnote{It is important to note that the distribution of exposure to ARMs includes values that surpass the regulatory limit of 66\%. However, this does not reflect the structure of the mortgage at its origination, but rather the structure at the time the data was collected. For instance, if a household initially took out a mortgage with one-third ARM and two-thirds indexed to the CPI, but later eliminated the CPI-fixed portion of the mortgage before the data collection, the remaining mortgage would be 100\% ARM.}

In 2013, the Bank Supervision Department of the Bank of Israel implemented a regulation limiting the maximum ARM ratio of a mortgage loan to two-thirds. This regulation also restricted the permissible variation within five years to no more than one-third. Consequently, only up to one-third of each mortgage could be directly tied to the Bank of Israel's policy rate. The remainder had to be allocated to either a fixed-rate format (at a minimum of one-third) or a five-year variable rate. In a notable policy shift in December 2020, the maximum portion of the mortgage that could fluctuate with the Bank of Israel policy rate was raised to two-thirds. This directive was swiftly enacted, taking effect in January 2021. The amendment underscores the dynamic nature of the Israeli mortgage market and the pivotal role of regulatory decisions in shaping its trajectory.\footnote{See Section \ref{subsection: prime regulation} for more details on the ARM ratio regulation change.}

\section{Data}
\label{Section: Data}

In this study, we use data from the Israeli Consumer Credit Register, which includes all consumer credit data for the entire population of borrowers in Israel. The Credit Register was established in 2016 as part of the "Credit Data Law" with the proclaimed goals of enhancing competition in the retail credit market through the sharing of credit information. The Credit Register is maintained by the Bank of Israel, and two private credit bureaus use this data to supply credit reports and scores about potential borrowers. All banks and credit card companies are required to report both their new and outstanding credit data on a monthly basis.\footnote{The Bank of Israel gathers and holds all the credit data, commonly referred to as the "Credit Register." This data is then transmitted to private credit bureaus, created following the law, which compute credit scores based on this information on a case-by-case basis. The Bank of Israel provides a website where consumers can obtain their credit history. Additional information regarding the Israeli Credit Data Register and its impact of consumer credit in Israel are available at: \url{https://www.creditdata.org.il/en} and \textcite{bank2023relationship}.} The Credit Register contains information on all consumer credit facilities, both new and outstanding, such as consumer loans, credit cards, credit lines, and mortgages, updated on a monthly basis.

For our empirical analysis, we extract a panel dataset of all mortgage borrowers in Israel, a group representing about 30\% of the households in Israel. We aggregate credit information for each household on a monthly basis, which includes details on their monthly outstanding mortgage debt. Specifically, our focus is on mortgage balances that are subject to an adjustable mortgage rate (ARM), which is directly linked to the Bank of Israel’s (BOI) prime rate. This means that the interest on these ARMs updates automatically and immediately following any changes in the BOI's policy rate. For each borrower, we calculate the 'ARM ratio' by dividing the balance of the ARM by the total outstanding mortgage balance. This ratio serves as our primary explanatory variable for assessing the exposure of mortgage borrowers to interest rate changes.  

We use credit card spending information to proxy for consumption patterns.\footnote{While we cannot observe cash withdrawals and cash payments, credit cards are by far the most popular means of payment in Israel.} Generally, most credit cards issued in Israel are, in fact, \textit{deferred debit} cards. There are no interest payments for using these cards, and the full outstanding balance is automatically withdrawn from the consumer's checking account once every month. While rollover credit cards such as those in the US, where consumers choose how much to pay every month and pay interest on the balance, exist, they are extremely uncommon.\footnote{One option for using the deferred debit for longer credit is to split a specific bill with a specific merchant over the course of several months (known in Hebrew as Tashlumim"). This method does include, in some cases, interest payments, and the split bill reduces the available credit limit during the payment schedule.} For each consumer, we aggregate separately all interest-bearing credit card balances and non-interest-bearing credit card balances, where the latter is defined as deferred debit" balance and is used as our main variable of interest for estimating consumption.

Our sample period spans from January 2021 through June 2023, encompassing the entirety of the recent interest rate increase period as well as a symmetric period of 15 months before the rates began to rise.\footnote{Our sample period begins after the Bank of Israel raised the maximum allowable portion of ARM from one-third to two-thirds (66\%).} For each household, we exclude months where there was no balance on any differed debit. Overall, our final baseline sample includes a little over 785 thousand households with roughly 21 million household-month observations. 

\begin{figure}[!htb]
\centering
\caption{Mortgage and deferred debt spending}
\includegraphics[width=1\textwidth]{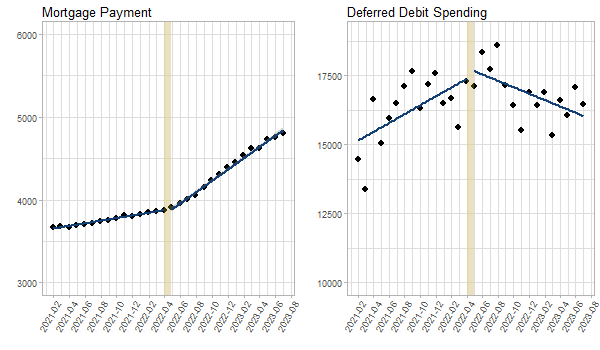}
\label{Fig motivation spending}
\floatfoot*{\textit{Notes}: This figure reports the average mortgage payment (left panel) and average deferred debt spending per household  from January 2021 to June 2023(right panel)}
\end{figure}

The motivation for our study is neatly summarized by Figure~\ref{Fig motivation spending}. The figure presents a comparative analysis of the average monthly mortgage payments and deferred debit spending for all mortgage borrowers in Israel at the household level. An upward trajectory is observed in mortgage payments, escalating from NIS 3,873 in March 2022 to NIS 4,808 by June 2023, marking a significant increase of approximately 25\% within a span of fifteen months.\footnote{There is a general upward trend in average mortgage payments, reflecting a persistent increase in the size of new mortgages against the backdrop of rising house prices throughout the period.} Simultaneously, the graph reveals a decline in credit card spending among the same group of mortgage borrowers, with the monthly average falling from NIS 17,300 in May 2022 to NIS 16,461 by June 2023, a decrease of approximately 5\% over the 13-month period.

Table~\ref{Table: Descriptive1} presents descriptive statistics for the entire sample. The term "Low ARM" represents observations where the borrower's ARM ratio is below the full sample median, and "High ARM" refers to observations where the ARM ratio is equal to or above the sample median (roughly 32\%.) 

\afterpage{%
\clearpage%
\thispagestyle{empty}%
\newgeometry{margin=1in}%
\begin{landscape}
\begin{table}[!htb] \centering 
  \caption{Descriptive statistics - All mortgagors split by ARM exposure} 
  \label{Table: Descriptive1} 
\footnotesize
\begin{tabularx}{\textwidth}{@{\extracolsep{6pt}}lYYYYYYYYYYYY} 
\\[-1.8ex]\hline 
\hline \\[-1.8ex] 
 & \multicolumn{4}{c}{All} & \multicolumn{4}{c}{Low ARM} & \multicolumn{4}{c}{High ARM} \\ 
\cline{2-5} \cline{6-9} \cline{10-13}
 & Mean & St. Dev & Q1 & Q3 & Mean & St. Dev & Q1 & Q3 & Mean & St. Dev & Q1 & Q3\\ 
\hline \\[-1.8ex]
Socio-Economic Index & $6.05$ & $2.11$ & $5$ & $8$ & $5.88$ & $2.04$ & $5$ & $7$ & $6.23$ & $2.16$ & $5$ & $8$ \\ 			
ARM ratio (\%) & $32$ & $24$ & $22$ & $38$ & $18$ & $15$ & $0$ & $30$ & $47$ & $22$ & $33$ & $54$ \\ 			
Mortg. Current Balance & $644,400$ & $616,199$ & $235,500$ & $883,900$ & $572,701$ & $582,689$ & $163,000$ & $811,500$ & $716,098$ & $639,995$ & $308,300$ & $949,000$ \\ 		
Mortg. Payment & $4,066$ & $2,796$ & $2,245$ & $5,155$ & $3,604$ & $2,625$ & $1,870$ & $4,650$ & $4,528$ & $2,885$ & $2,665$ & $5,610$ \\ 		
Consumer Loan (\%)  & $61$ & $49$ & $0$ & $100$ & $64$ & $48$ & $0$ & $100$ & $59$ & $49$ & $0$ & $100$ \\
Consumer Loan (Balance) & $86,871$ & $140,817$ & $0$ & $115,600$ & $87,851$ & $137,363$ & $0$ & $119,250$ & $85,892$ & $144,184$ & $0$ & $111,450$ \\ 
Overdraft (\%) & $39$ & $48$ & $0$ & $100$ & $42$ & $49$ & $0$ & $100$ & $35$ & $48$ & $0$ & $100$ \\ 		 	
Overdraft (Balance) & $7,548$ & $15,840$ & $0$ & $7,850$ & $8,172$ & $16,099$ & $0$ & $9,780$ & $6,925$ & $15,552$ & $0$ & $5,650$\\
Credit Card Loan (\%) & $36$ & $48$ & $0$ & $100$ & $40$ & $49$ & $0$ & $100$ & $32$ & $47$ & $0$ & $100$ \\
Credit Card Loan (Balance) & $8,226$ & $24,820$ & $0$ & $1,975$ & $9,151$ & $25,807$ & $0$ & $2,890$ & $7,302$ & $23,757$ & $0$ & $1,200$ \\				
Deferred Debit (balance)  & $16,578$ & $14,575$ & $6,385$ & $22,180$ & $15,040$ & $13,692$ & $5,540$ & $20,150$ & $18,117$ & $15,255$ & $7,400$ & $24,150$ \\ 
Arrears (\%) & $1$ & $10$ & $0$ & $0$ & $2$ & $12$ & $0$ & $0$ & $1$ & $8$ & $0$ & $0$ \\
Observations & \multicolumn{4}{c}{21,287,084} & \multicolumn{4}{c}{10,643,527}& \multicolumn{4}{c}{10,643,557}\\
Households & \multicolumn{4}{c}{78,5562} & \multicolumn{4}{c}{39,7705}& \multicolumn{4}{c}{38,7857}\\ 
\hline \\[-1.8ex]  
\multicolumn{13}{p{.95\textwidth}}{\textit{Notes:} This table presents the descriptive statistics for the final sample of mortgage households. All observations are recorded at the household-month level. Columns 1 through 4 detail the full sample. Columns 5 through 8 focus on households whose ARM ratio – the share of the mortgage's unpaid current balance that is directly linked to the Bank of Israel's interest rate – falls below the median of the distribution (roughly 32\%.) Conversely, Columns 9 through 12 focus on borrowers with an ARM ratio above the sample median. The data covers the sample period from January 2021 to June 2023. For detailed information on the construction of the sample and the variables, refer to Section \ref{Section: Data}. The table displays the mean, standard deviation, and the 25th (Q1) and 75th (Q3) percentiles for each variable.  \par} \\ 
\end{tabularx} 
\end{table} 

\vfill 
\begin{center}
\thepage
\end{center}

\end{landscape}
\clearpage%
\restoregeometry%
}


The socio-economic indicator is based on the municipality where the borrower resides. The Israeli Central Bureau of Statistics provides a socioeconomic index ranging from 1 to 10 for each local council or municipality, where one represents the poorest socioeconomic conditions and ten the highest. The mortgage ARM ratio is our main explanatory variable and is defined as the share of the borrower monthly current balance that is directly linked to the BOI. Our main dependent variable is the borrower deferred debit monthly balance from all credit cards. 

We also control for three additional type of consumer debt. Overdraft debt which is a credit line that banks grant their clients on their checking accounts from which they can withdraw funds up to some limit, and Consumer loans which are all other term loans which consumer use that include car loans and other general purpose loans. Credit card loans refer to the aggregate balance from all rollover credit cards and balances that carry interest on regular cards. 

\textcolor{black} {Table \ref{Table: Descriptive1} provides valuable insights into the financial characteristics of Israeli mortgage borrowers during our study period from January 2021 to June 2023. The data reveals notable differences between the High ARM and Low ARM groups, categorized based on their adjustable-rate mortgage (ARM) exposure. The High ARM group, with an average ARM ratio of 47\%, exhibits a substantially higher average mortgage balance of NIS 716,098 compared to NIS 572,701 for the Low ARM group, which has an average ARM ratio of 18\%. Consequently, the High ARM group faces higher monthly mortgage payments, averaging NIS 4,528, while the Low ARM group's average payment stands at NIS 3,604. These figures underscore the greater financial exposure to interest rate fluctuations faced by households with a higher proportion of adjustable-rate mortgages.}

\textcolor{black}{Furthermore, the High ARM group appears to have a slightly higher socio-economic index, with an average of 6.23 compared to 5.88 for the Low ARM group, suggesting a potential correlation between economic status and the propensity to take on adjustable-rate mortgages. The data also reveals differences in consumer credit utilization between the two groups. While a larger proportion of the Low ARM group holds consumer loans (64\% vs. 59\%) and overdrafts (42\% vs. 35\%), the High ARM group maintains higher average balances in both categories (NIS 85,892 vs. NIS 87,851 for consumer loans and NIS 6,925 vs. NIS 8,172 for overdrafts).}

Table \ref{Table: Descriptive1a} shows the descriptive statistics for the datset when split by observation in the period before the rise in interest rates (January 2021-March 2022) and the period after (April 2022-June 2023). The table shows that the pre and post samples are overall similar which suggests that there was no large shifts in borrower composition during the period.

\begin{table}[!htb] \centering 
  \caption{Descriptive statistics - All mortgagors split by period} 
  \label{Table: Descriptive1a} 
\scriptsize
\begin{tabularx}{.95\textwidth}{@{\extracolsep{6pt}}lYYYYYYYY} 
\\[-1.8ex]\hline 
\hline \\[-1.8ex] 
 & \multicolumn{4}{c}{January 2021-March 2022} & \multicolumn{4}{c}{April 2022-June 2023} \\ 
\cline{2-5} \cline{6-9}
 & Mean & St. Dev & Q1 & Q3 & Mean & St. Dev & Q1 & Q3\\ 
\hline \\[-1.8ex]
Socio-Economic Index & $6.06$ & $2.10$ & $5$ & $8$ & $6.04$ & $2.12$ & $5$ & $8$ \\ 
ARM Ratio (\%) & $32$ & $24$ & $23$ & $37$ & $32$ & $24$ & $22$ & $39$ \\ 
Mortg. Current Balance & $611,332$ & $579,545$ & $222,400$ & $844,000$ & $676,980$ & $648,648$ & $249,700$ & $922,950$ \\ 
Mortg. Payment & $3,765$ & $2,510$ & $2,080$ & $4,780$ & $4,363$ & $2,947$ & $2,440$ & $5,525$ \\ 	
Consumer Loan (\%) & $62$ & $49$ & $0$ & $100$ & $61$ & $49$ & $0$ & $100$ \\
Consumer Loan (Balance)  & $84,053$ & $136,724$ & $0$ & $111,250$ & $89,649$ & $144,685$ & $0$ & $120,000$ \\					
Overdraft (\%) & $38$ & $49$ & $0$ & $100$ & $39$ & $49$ & $0$ & $100$ \\
Overdraft (Balance)& $7,266$ & $15,453$ & $0$ & $7,350$ & $7,828$ & $16	,209$ & $0$ & $8,350$ \\  		
Credit Card Loan (\%) & $36$ & $48$ & $0$ & $100$ & $37$ & $48$ & $0$ & $100$ \\
Credit Card Loan (Balance)& $7,386$ & $22,779$ & $0$ & $1,755$ & $9,056$ & $26,655$ & $0$ & $2,200$ \\ 
Differed Debit (balance) & $16,292$ & $14	,299$ & $6,350$ & $21,715$  & $16,860$ & $14,839$ & $6,410$ & $22,650$ \\ 	 		
Arrears (\%)  & $1$ & $11$ & $0$ & $0$ & $1$ & $10$ & $0$ & $0$ \\
Observations & \multicolumn{4}{c}{10,564,348} & \multicolumn{4}{c}{10,722,736}\\	
Households & \multicolumn{4}{c}{785,562} & \multicolumn{4}{c}{785,562}\\	
\hline \\[-1.8ex]  
\multicolumn{9}{p{.95\textwidth}}{\textit{Notes:} This table presents the descriptive statistics for the household with mortgages dataset. All observations are recorded at the household-month level. Columns 1 through 4 detail observations in the pre BOI policy hike - January 2021 through March 2022. Conversely, Columns 5 to 8 present observation the the post period - April 2022 through June 2023. For detailed information on the construction of the sample and the variables, refer to Section \ref{Section: Data}. The table displays the mean, standard deviation, and the 25th (Q1) and 75th (Q3) percentiles for each variable.  \par} \\ 
\end{tabularx} 
\end{table} 

For the estimation around the ARM regulation change, we focus exclusively on borrowers whose mortgages were originated around the time of the regulatory shift. Specifically, we include only those borrowers with mortgages originating from October 2020 to April 2021. 
Borrowers with mortgages initiated before mid-January 2021 constitute the control group, while those with mortgages starting from mid-January 2021 are deemed the treated group.\footnote{Mortgage starting dates are reported biweekly in the credit registry.} Our examination then centers on the consumption behavior of these two groups from December 2021 through June 2023.

\begin{table}[!htb] \centering 
  \caption{Descriptive statistics - Mortgages originated around January 2021} 
  \label{Table: Descriptive2} 
\scriptsize
\begin{tabularx}{.95\textwidth}{@{\extracolsep{6pt}}lYYYYYYYY} 
\\[-1.8ex]\hline 
\hline \\[-1.8ex] 
 & \multicolumn{4}{c}{Before} & \multicolumn{4}{c}{After} \\ 
\cline{2-5} \cline{6-9}
 & Mean & St. Dev & Q1 & Q3 & Mean & St. Dev & Q1 & Q3\\ 
\hline \\[-1.8ex]
\multicolumn{9}{l}{\textbf{Panel A. All Households}}\\[1.4ex]  
Socio-Economic Index  & $5.96$ & $2.14$ & $5$ & $8$ & $6.04$ & $2.11$ & $5$ & $8$ \\ 
ARM Ratio (\%) & $32$ & $12$ & $30$ & $33$ & $36$ & $16$ & $32$ & $39$ \\ 		
Mortg. Current Balance & $910,881$ & $633,835$ & $564,808$ & $1,111,000$ & $942,660$ & $617,044$ & $582,000$ & $1,158,900$ \\ 
Mortg. Payment & $4,757$ & $2,545$ & $3,140$ & $5	,770$ & $4,858$ & $2,593$ & $3,175$ & $5,910$ \\
Consumer Loan (\%)& $62$ & $49$ & $0$ & $100$ & $60$ & $49$ & $0$ & $100$ \\  	
Consumer Loan (Balance) & $92,243$ & $141,722$ & $0$ & $126,800$ & $89,686$ & $140,994$ & $0$ & $122,800$ \\ 		
Overdraft (\%) & $38$ & $49$ & $0$ & $100$ & $38$ & $48$ & $0$ & $100$ \\ 						
Overdraft (Balance) & $7,170$ & $15,232$ & $0$ & $7,280$ & $6,909$ & $14,899$ & $0$ & $6,600$ \\ 		
Credit Card Loan (\%) & $36$ & $48$ & $0$ & $100$ & $35$ & $48$ & $0$ & $100$ \\
Credit Card Loan (Balance) & $8,870$ & $26,496$ & $0$ & $1,850$ & $8	,596$ & $25,955$ & $0$ & $1,795$ \\ 		
Differed Debit (Balance) & $16,022$ & $14	,189$ & $6,300$ & $21,220$ & $16,258$ & $14,382$ & $6,359$ & $21,450$ \\ 	
Arrears (\%) & $1$ & $8$ & $0$ & $0$ & $0.5$ & $7$ & $0$ & $0$ \\	
Observations & \multicolumn{4}{c}{394,295} & \multicolumn{4}{c}{432,112}\\
Households & \multicolumn{4}{c}{21,486} & \multicolumn{4}{c}{23,500}\\ \\ 
\multicolumn{9}{l}{\textbf{Panel B. Top Two Terciles}}\\[1.4ex] 
Socio-Economic Index  & $6.07$ & $2.16$ & $5$ & $8$ & $6.13$ & $2.15$ & $5$ & $8$ \\
ARM Ratio (\%) & $35$ & $12$ & $32$ & $34$ & $41$ & $15$ & $33$ & $49$ \\
Mortg. Current Balance & $944,171$ & $610	,435$ & $592,000$ & $1,151,000$ & $977,653$ & $638,311$ & $609,000$ & $1,190,200$ \\  
Mortg. Payment & $5,026$ & $2,542$ & $3,420$ & $6,015$ & $5,153$ & $2,631$ & $3,455$ & $6,200$ \\ 	
Consumer Loan (\%) & $0.60$ & $0.49$ & $0$ & $1$ & $58$ & $49$ & $0$ & $100$ \\ 			
Consumer Loan (Balance) & $88,066$ & $139	,533$ & $0$ & $118,850$ & $84,520$ & $137,890$ & $0$ & $113,000$ \\		
Overdraft (\%) & $36$ & $48$ & $0$ & $100$ & $34$ & $47$ & $0$ & $100$ \\ 	 					
Overdraft (Balance) & $6,548$ & $14,618$ & $0$ & $5,650$ & $6,166$ & $14,201$ & $0$ & $4,575$ \\		
Credit Card Loan (\%) & $32$ & $47$ & $0$ & $100$ & $32$ & $46$ & $0$ & $100$ \\
Credit Card Loan (Balance) & $7,497$ & $24,417$ & $0$ & $1,220$  & $7,092$ & $23,544$ & $0$ & $1,090$ \\ 			
Differed Debit (Balance) & $16,528$ & $14	,355$ & $6,635$ & $21,900$ & $16,767$ & $14,622$ & $6,655$ & $22,165$ \\ 	 	
Arrears (\%) & $1$ & $7$ & $0$ & $0$ & $0.4$ & $6$ & $0$ & $0$ \\	
Observations & \multicolumn{4}{c}{262,858} & \multicolumn{4}{c}{288,074}\\
Households & \multicolumn{4}{c}{14,324} & \multicolumn{4}{c}{15,136}\\ \\[-1.8ex] 
\hline \\[-1.8ex]  
\multicolumn{9}{p{.95\textwidth}}{\textit{Notes:} This table presents the descriptive statistics of mortgage borrowers in Israel. All observations are at the individual borrower-month observation level. Columns 1 through 4 present borrowers with mortgages originated before the regulatory change increasing the maximum variable rate portion of a mortgage (October 2020 - mid January 2021). Columns 5-8 present borrowers with mortgages originated after the regulatory change (mid January 2021 - April 2021). The sample period is January 2021 to June 2023. For details on the construction of the sample and the variables, see Section~\ref{Section: Data}. Mean, standard deviation, 25th and 75th percentile are presented for each variable.} \\ 
\end{tabularx} 
\end{table}

Table \ref{Table: Descriptive2} presents descriptive statistics of mortgage borrowers in Israel who originated mortgages around January 2021, specifically before and after the regulatory change that increased the maximum variable rate portion allowed in a mortgage. Panel A uses the full sample while panel B excludes for each group the lower ARM ratio tercile, keeping only the households that were likely more constraint/impacted by the regulation change. Overall,the comparison reveals that the borrowers who obtained mortgages post-regulation change share many characteristics with those who obtained them before the change. They generally have comparable socio-economic statuses, and exhibit consistent borrowing and debt patterns, with only minor variations in certain financial metrics. Furthermore, the average current mortgage balances and monthly mortgage payments, while higher in the post-change group, follow a consistent trend, suggesting a stable mortgage market environment. In terms of additional debts, both groups show a propensity to hold consumer and credit card loans. Overall, the two groups seem quite similar with an important distinction in the mortgage ARM ratio where the treated group mean is four percent point higher for the full sample and six percentage points higher of the upper two terciles of each group distribution. For the 75$^{th}$ percentile the mean ARM ratio is nine percentage points higher for all mortgages and 15 percentge points higher for the upper terciles.

\section{Empirical Framework}
\label{Section: empirical}

\subsection{All Mortgages}
Our study employs a unique aspect of Israeli mortgages to understand the impact of policy rate changes on consumption expenditures. We focus on the variation in mortgage borrowers' direct exposure to changes in the Bank of Israel rate. This approach enables us to explore the relationship between mortgage structure and the variation in deferred debit spending following these policy changes.

The context of our research is crucial: a significant proportion of Israeli loans are ARMs, intrinsically linked to the Bank of Israel interest rate. Our analysis contrasts periods of relatively stable interest rates with a period marked by a rapid and largely unforeseen increase in these rates. We hypothesize that, had the interest rates consistently remained low, the differences in expenditure between  mortgage borrowers with high and low interest rate exposure would have remained largely unchanged. This assumption is key to isolating the impact of fluctuating interest rates on household spending, thus providing clearer insights into economic behaviors under variable interest rate conditions.

To empirically test our hypothesis, we apply a regression model that accounts for individual and time fixed effects, the share of ARM balance, and various other controls. Our baseline model, which incorporates these elements, is designed to estimate the relationship between spending, ARM exposure, and other relevant variables:

\begin{align}
\label{equ1}
\log(\text{DeferredDebit})_{i, t} = \beta \left(\text{ARMRatio}_{i}\times\text{Post}_t\right) 
+ \theta \left(X_{i} \times \text{Post}_t\right) + \alpha_i + \gamma_t + \epsilon_{it}
\end{align}


where $\log(\text{DeferredDebit})_{i, t}$ represents the natural logarithm of the end-of-the-month total deferred debit spending by household $i$ during month $t$, $\text{ARMRatio}_{i}$ denotes the average ratio of the unpaid current balance that is directly linked to the prime rate from the total outstanding mortgage balance, in the pre period. The equation incorporates $\alpha_i$ and $\gamma_t$ to represent individual and time fixed effects, respectively. The dummy variable $\text{Post}_t$ takes the value of one for the period from April 2022 through June 2023, a time when the policy rate was continuously increasing. $X_{i}$ refers to household level control variables s that capture additional forms of pre-existing non-mortgage debt such as consumer loans, overdrafts, and credit card loans.

This specification enables a direct estimation of the relationship between mortgage borrowers' exposure to the Bank of Israel interest rate and their spending habits. Here, our main variable of interest is $\beta$ which captures the change in spending behavior during the period of escalating interest rates, from April 2022 to June 2023.

On the one hand, when facing a rapid increase in mortgage payments, borrowers may attempt to mitigate the impact on their disposable income by depleting savings and/or incurring additional debt. This approach could potentially buffer the liquidity shock. On the other hand, exposure to other types of debt and changes in their monthly payments can affect disposable income and, consequently, household consumption. To account for this, we include in our model interactions between the post dummy and other types of households' debt. Specifically, a household without consumer loans, overdrafts, or credit card debt might have more options to smooth out the liquidity shock, at least in the short term, compared to borrowers who have already utilized all other debt avenues. In our baseline specification, we use a dummy variable to indicate whether the household possesses any of these debt instruments in six month period before rates started to increase. Specifically, for consumer loans we use a dummy equal to one for those households who had an outstanding balance in the six months before April 2022 and for overdraft and credit card debt we use a dummy equal to one for households with such debt in at least three of the six months.\footnote{Our aim is to control for credit constraint households, while for consumer loans that implies pre-existing outstanding debt, households which move in and out of overdraft/credit card debt are also likely experiencing liquidity shortage and therefore we use at least three of the six months as an indication for credit constraints and not the full six months.}   
The robustness of our results is further verified in in Section \ref{subsection: robustness all}, where we substitute the dummy variable with the average unpaid balances of these debts in the six months before April 2024.

We also examine the dynamic relationship between the ARM ratio and borrower-level dependent variables by replacing $\text{Post}$ with a series of dummy variables. These variables span 15 months before and 15 months after the start of the interest rate increase. This dynamic specification allows us to investigate whether borrowers with different ARM ratio values were experiencing distinct pre-existing trends in deferred credit card spending prior to the period when the rate increased.

One might question whether using the actual change in mortgage payment would be more appropriate than using the pre period average ARM ratio. However, since the actual change in payment is endogenously determined, employing it as the main independent variable could yield biased results. For instance, a borrower with a high ARM Ratio might, when faced with a significant increase in the policy rate, attempt to mitigate the payment increase through several methods. These could include cutting spending or using savings to pay down parts of the mortgage, potentially resulting in an actual \textit{reduction} in the mortgage payment alongside a decrease in debit card spending. Therefore, we use the ARM ratio of the mortgage borrower in the pre period, as it represents a predetermined \textit{exposure} to interest rate hikes, i.e., the potential decrease in disposable income. In Section \ref{subsection: robustness all}, we demonstrate that our results are also robust when using the actual amount of the mortgage payment. 
Additionally, in Section \ref{subsection: robustness all}, we confirm that our main findings are consistent across a wide array of additional robustness tests. 

It is also important to note why we use only the ARM track to proxy for borrowers' exposure to interest rates. As mentioned in Section \ref{Section: mort Isr}, borrowers' mortgage payments may also be impacted by the inflation rate through the CPI-linked tracks and interest rates through the 5-year variable (Medium-Fixed) rate track. First, CPI-linked tracks do not have a large immediate impact on mortgage payments. For example, a 25-year, 100,000 NIS mortgage track that is fixed rate at 3\% and linked to the CPI will pay 474 NIS in the first month. A two percent annual inflation rate (within the BOI inflation target range) will then increase the monthly payment every month, on average, by one NIS. Thus, even an increase of inflation to 5.2\% (the peak point in 2022 and over the last two decades in Israel) will still imply a very modest immediate monthly payment increase. It is, of course, crucial to note that the increase in payments in the CPI-linked track could have a very substantial impact on the overall cost of the mortgage due to the compounding effect of inflation; however, the impact in terms of borrowers' liquidity (cash flow) is relatively small.\footnote{For comparison, a 50 basis point increase in the policy rate will, on average, increase the monthly payment on a 100,000 NIS mortgage in the ARM track by around 30 NIS.}

Regarding the Medium-Fixed tracks, as the interest on these tracks is adjusted every five years, during our sample period (30 months), only a small share of the mortgage payment adjustments will be due to this track.\footnote{Of the Medium-Fixed tracks, approximately half will have their anchor interest rate updated once during the sample period, and of those, half will be in the period before the interest rate hikes. Note that the 5-year variable rate tracks are also relatively less popular and account for 23\% of the total balance. Thus, payments related to only about 6\% of the total mortgage balances changed in the post period due to this track.} In the robustness test section, we show that the results are not impacted when adding additional mortgage structure controls including the share of medium-fixed and linked mortgages.

\subsection{ARM Regulation Change}
\label{subsection: prime regulation}
Thus far, our empirical evidence is predicated on the assumption that the ARM ratio of each mortgage borrower is largely exogenous to time-varying factors that could influence the relationship between policy rate changes and borrower consumption. This means that the factors determining the variation across borrowers are not the same factors driving the results; hence, there is no additional selection or omitted variable bias causing some borrowers to have mortgages with more ARM exposure while also being more sensitive to high inflation or rising interest rates. If this assumption is incorrect, our results might be biased due to an obvious selection bias of consumers opting for variable rates over fixed-rate mortgages.

To address this potential selection bias, we employ a difference-in-differences (DiD) regression methodology, taking advantage of an unanticipated regulatory change that affected mortgage borrowers' exposure to interest rates. As detailed in Section \ref{Section: mort Isr}, amidst the COVID crisis, in January 15, 2021, the Bank of Israel's Banking Supervision Department increased the maximum limit of the ARM track from 33 percent to 66 percent. This regulatory amendment was unexpected and garnered significant media and public attention. At the time, the combination of high housing demand and the expectation that the "Low for long" interest rate environment would persist induced a strong demand for ARM tracks due to their low current cost. Consequently, the regulation, which was binding for many borrowers, led to an almost immediate jump in the average prime rate for new mortgages.

\begin{figure}[!htb]
\centering
\includegraphics[width=0.9\textwidth]{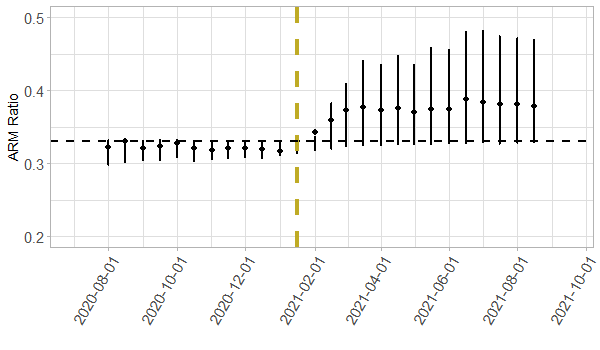}
\caption{ARM Ratio and the ARM Regulation Change}
\label{Fig motivation prime}
\floatfoot*{\textit{Notes}: This figure illustrates the average ARM ratio of mortgages with first to third quartile ranges, categorized by the biweekly origination date of the mortgage. The ARM ratio is defined as the proportion of the unpaid current balance directly linked to the prime rate in relation to the total mortgage balance. Horzontal black dashed line is at the 33\%, representing the maximum ARM share before January 15$^{th}$ 2021.}
\end{figure}

Figure \ref{Fig motivation prime} illustrates the average ARM ratio by mortgage origination date with first to third quartile ranges. We observe a distinct increase from an average ARM ratio of 32.2\% before the regulatory change to 37\% afterwards, along with higher variability (as indicated by the confidence bars). It is noteworthy that while the regulation was binding, the new limit was not fully utilized; that is, the ratio did not increase all the way up to 66\%. This suggests that while borrowers adjusted to the new regulation, most did not do so to the full extent allowed. 

Figure \ref{Fig motivation prime} also demonstrates our rationale for selecting a three-month window around the regulation change for the baseline estimation. The identification strategy relies on comparing mortgages originated under similar market conditions. Ideally, we would prefer the narrowest possible time window to ensure the most comparable circumstances. However, the figure clearly indicates that it took approximately two to three months for the average ARM share to stabilize at a new market equilibrium following the regulation change. Therefore, we choose a three-month window as it represents the shortest time frame that still captures a clear shift in the ARM level, balancing the need for both comparability and a sufficient sample size. This choice allows us to effectively isolate the impact of the regulation change on borrowers' ARM exposure while ensuring the robustness of our results. In the robustness section, we further demonstrate that our findings remain consistent when using alternative two-month and four-month windows around the regulation change.

\begin{figure}[!htb]
\centering
\caption{ARM Ratio Density Around Regulation Change}
\includegraphics[width=0.8\textwidth]{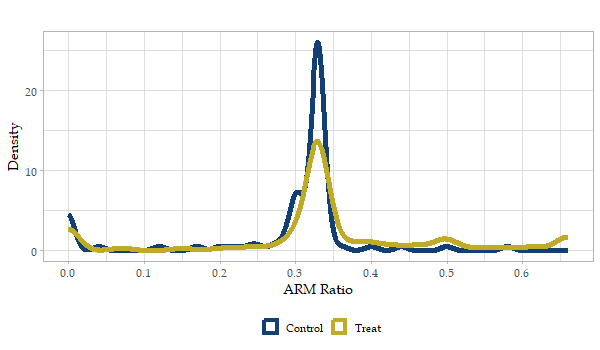}
\label{Fig mort density regulation}
\floatfoot*{\textit{Notes}: This figure presents the density of the ARM ratio from each mortgage current balance for mortgages that originated between October 2020 to mid January 2020 (control) and from mid January 2021 to April 2021 (treatment).}
\end{figure}

\textcolor{black}{Figure \ref{Fig mort density regulation} presents the distribution of mortgage borrowers' ARM ratios for those whose mortgages were originated around the time of the regulatory change indicates a notable pattern. Mortgages that originated before the regulatory change are highly concentrated at the regulatory limit. Specifically, 45\% of borrowers whose mortgages were initiated between August 2020 and November 2020 had, as of February 2022, a ARM ratio of exactly 33\% (with 60\% of these borrowers' ARM ratios ranging between 32-34\%). In contrast, borrowers with mortgages originating following the regulatory change show a more dispersed distribution with a lower proportion at the 33\% ARM ratio, although this rate remained the most common. This pattern indicates a possible shift in borrower behavior or financial institution strategies following the regulatory amendment.}

Building on these observations, we employ the following DiD design. In our baseline estimation the treatment group consists of households whose mortgages originated in following the regulatory change. The control group comprises borrowers with mortgages from the period preceding the change:
\begin{align}
\label{equ2}
\log(\text{DeferredDebit})_{i, t} = \beta\left(\text{Post}_t \times treat_{i} \right)
+ \theta\left(\text{Post}_t \times X_{i}\right) + \alpha_i + \gamma_t + \epsilon_i
\end{align}

where $treat_i$ is an indicator set to one for the treatment group, and all other variables are as defined in Equation~(\ref{equ1}).\footnote{\textcolor{black}{For the DiD specification, we begin the sample period in December 2021 to ensure that all mortgages in the sample have been fully drawn down.}} Since for some borrowers, both in the control and the treatment groups, a 33\% ARM ratio limit was not necessarily a binding constraint, we also estimate the DiD specification after restricting both samples to those borrowers whose average ARM ratio in the period before interest rates began to rise (January 2022 through March 2022) was in the upper two terciles of the distribution in each group. This approach effectively narrows our focus to borrowers in the control group who likely would have chosen a higher ARM ratio had the opportunity been available to them.

\section{Main Results}
\label{Section: Results}

In this section, we present a comprehensive analysis of the impact of interest rate changes on borrower consumption in Israel, specifically focusing on mortgage holders. Utilizing a robust dataset, we apply Equation~\eqref{equ1} to explore how adjustments in mortgage interest rates, particularly those linked to the BOI rate, influence the spending behaviors of consumers as observed through their deferred debit spending. Our approach is rooted in a detailed examination of the ARM ratio and its interaction with the period following the Bank of Israel's initiation of a rate hike campaign in April 2022, identified as "Post" in our study.

\subsection{All Mortgages}
\label{subsection: results all}

Table \ref{Table: main results} shows our baseline estimates for the effect of interest rate changes on borrowers' consumption, based on applying Equation~\eqref{equ1} to the log of deferred debit card spending as the dependent variable. "ARM ratio" is the average fraction of the unpaid mortgage balance that is pegged to the prime rate compared to the total balance of the mortgage in the "Pre" period. "Post" is a binary dummy variable that takes the value of 1 for the period from April 2022 onward, the month following which the Bank of Israel started its rate hike campaign. The coefficient of the interaction between the ARM ratio and Post is our main focus as it measures how a higher share of adjustable-rate mortgage relates to the impact of the monetary contraction period on the (log) level of deferred debit (our measure of consumption).

The interaction effect in the regression we run is given in basis points per the average change in the interest rate during the hiking period that is captured by "Post" (April 2022 - June 2023.) The first column of the table includes "ARM ratio" and its interaction with "Post." In column two, we add additional interactions between the "Post" dummy and other financial characteristics of the household - Consumer Loans, Overdrafts, and Credit Card. All regressions include two-way fixed effects at the level of the borrower-month.

The interaction between ARM ratio and $\text{Post}$ reveals that during the period of rising interest rates, mortgagors with higher ARM ratios reduced their consumption more significantly than those with lower ARM ratios. The coefficient implies that for every 1 percentage point higher ARM ratio, consumption decreased by 3.6-4.5 basis points more in the post period, during which the monetary policy rate tightened by 4.65 percentage points. To put this in perspective, the results imply that, all else being equal, a household with a 100\% ARM mortgage would have reduced consumption spending by approximately 3.6-4.5 percent more relative to a household with a 100\% fixed-rate mortgage (FRM).

The coefficients for the interaction of additional credit control variables with "Post" in column (2) are uniformly negative and statistically significant.\footnote{In this specification, the credit control variables are dummies that take the value of 1 if the household has this specific type of credit and zero otherwise.} This indicates that consumer loans, overdrafts, and credit cards follow the trends observed in a higher interest rate environment. It's important to note, however, that while our primary focus in this paper is on mortgage payments, the evidence presented for other financial variables related to the "cash-flow channel" suggests a broader effect on consumption through various paths, not just through mortgage payments. Our emphasis on mortgage payments in this study is mainly due to the data's capacity to leverage exogenous variation, allowing us to establish a credible causal link between interest rates and consumption.\footnote{To conserve space, we present only the results of the ARM ratio in the rest of this paper. The results for the other debt controls and interactions are in line with the baseline estimations and are available upon request.}

\begin{table}[!htb] \centering 
  \caption{Baseline estimations} 
  \label{Table: main results} 
\footnotesize
\begin{tabularx}{.9\textwidth}{@{\extracolsep{6pt}}lYY}
\\[-1.8ex]\hline 
\hline \\[-1.8ex] 
 & \multicolumn{2}{c}{Log(Deferred debit)} \\ 
\cline{2-3} 
\\[-1.8ex] 
\\[-1.8ex] & (1) & (2) \\ 
\hline \\[-1.8ex]

ARM ratio$\times$Post  & $-$0.036$^{***}$ & $-$0.045$^{***}$ \\ 	
  & (0.003) & (0.003) \\ 	
Consumer loans$\times$Post &  & $-$0.015$^{***}$ \\ 	
  &  & (0.001) \\ 	
Overdraft$\times$Post &  & $-$0.009$^{***}$ \\ 	
  &  & (0.001) \\ 	
Credit card$\times$Post &  & $-$0.047$^{***}$ \\ 	
  &  & (0.001) \\ 	
 \hline \\[-1.8ex] 
Household f.e & Y & Y \\ 	
Time f.e & Y & Y \\ 	
Credit Controls Interaction & N & Y \\ 				
Observations & 21,287,084 & 21,287,084 \\ 
R$^{2}$ & 0.762 & 0.762 \\ 				
Adjusted R$^{2}$ & 0.753 & 0.753 \\ 					
\hline 
\hline \\[-1.8ex] 
\multicolumn{3}{p{0.9\textwidth}}{\textit{Notes:} This table reports the coefficient estimates of Equation (\ref{equ1}) for the full sample of households with mortgages borrowers in Israel from January 2021 through June 2023. Columns represent different combinations of other borrower credit related dummy control variables. Standard errors clustered by borrower are reported in parentheses. $^{*}$p$<$0.1; $^{**}$p$<$0.05; $^{***}$p$<$0.01} 
\end{tabularx} 
\end{table} 

\begin{figure}[!htb]
\centering
\includegraphics[width=0.9\textwidth]{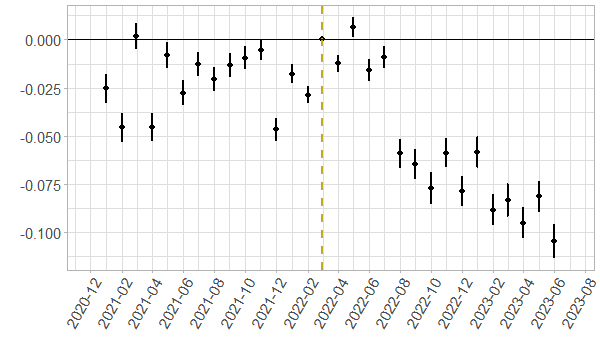}
\caption{Dynamic Impact of Mortgage Borrowers' Prime Rate on Monthly Deferred Debit Card Spending}
\label{Fig dynamic main}
\floatfoot*{\textit{Notes}: This figure reports the dynamic relationship between mortgage borrowers' ARM ratio and the logarithm of monthly deferred debit card spending. The coefficient estimates presented come from the estimation of Equation~\eqref{equ1}, which includes interactions between "ARM ratio" and monthly dummies for each month from February 2021 to June 2023, displayed with 90\% confidence bands. The coefficient for April 2022, the month when the Bank of Israel started hiking rates, is normalized to zero.}
\end{figure}

The temporal dynamics of how mortgage borrowers' ARM ratio impacts deferred debit card spending appear in Figure \ref{Fig dynamic main}. 
The figure shows estimates derived from the interactions between ARM ratio and monthly dummies in Equation~\eqref{equ1}'s estimation, presented with 90\% confidence bands. The benchmark for these estimates is March 2022's coefficient which is normalized to zero which coincides with the last month before the Bank of Israel starting its interest rate hike campaign.

The dynamic specification reveals that before April 2022, the date of the Bank of Israel's first rate hike, there was no clear and significant pattern in the interaction coefficient between "ARM Ratio" and the period dummy. However, after the first rate hike, a significant drop in the interaction coefficient becomes evident, especially from August 2022 onwards, after the BOI started aggressively hiking rates.\footnote{After three relatively mild increases in April, May, and July that brought the BOI policy rate to 1.25\%, the BOI increased rates by 75 basis points in August, which was, at that point, the largest single interest rate change in over two decades.} This suggests a persistent and long-lasting effect of the new, higher interest rate environment on spending patterns. Most importantly, the results of the dynamic specification do not enforce a specific cutoff date, as we do with "Post" in the static specification, and nonetheless, show that our selection of the cutoff date as the date on which the Bank of Israel began raising rates is justified. Importantly, the dynamic specification also reveals that by June 2023, the difference in consumption response was around ten basis points for every 1\% higher ARM ratio. This is consistent with the overall consistent interest rate increases during the post period, which corresponded to an overall continuous drop in consumption, of which the non-dynamic specification only captured the average.

\subsubsection{Quantifying the Economic Impact and MPC}
To assess the economic significance of our findings, we provide a back-of-the-envelope calculation of the mortgage cash flow channel's impact on total consumption and the marginal propensity to consume (MPC).

According to Table 2, the average spending of mortgaged households in the pre-hike period was 16,292 NIS, and their average ARM ratio was 32\%. Using the estimated coefficient from column 2 in Table 4, we calculate that the additional reduction in consumption for these households in the post-hike period was approximately 235 NIS per month ($16,292 \times 0.32 \times 0.045$), with respect to zero ARM. The average mortgage payment increased from 3,765 NIS in the pre-period to 4,363 NIS in the post-period, a difference of 598 NIS (see Table~\ref{Table: Descriptive1a}). Given these figures, the implied average MPC is 0.39 ($235 \div 598$).\footnote{This calculation assumes that an increase in mortgage payments is equivalent to a reduction in disposable income.}

In Israel, about 30\% of households have a mortgage. Since the average mortgage in Israel has an ARM ratio of 0.32, the reduction in mortgage borrowers consumption during the monetary policy tightening spell due to the mortgage cash-flow channel, compared to a counterfactual of zero ARM ratio, is 1.44\% ($0.32 \times 0.045$). Assuming that mortgaged households' consumption aligns with their population share, the estimated 1.44\% reduction in their monthly consumption due to the cash flow channel translates to a 0.43\% reduction in total private consumption during the post-hike period ($1.4\% \times 0.3$).

Building on these findings, we can extend our analysis to consider the full impact of the monetary tightening cycle on consumption. Notably, Figure~\ref{Fig dynamic main} shows that the consumption decline was twice as large with respect to our baseline results by the end of the monetary tightening cycle. Consequently, our results suggest that following the 4.65 percentage point policy rate hike, the mortgage cash flow channel accounted for an additional 2.88\% ($1.44\% \times 2$) reduction in mortgaged households' consumption and an overall reduction of approximately 0.86\% in total private consumption in Israel ($2.88\% \times 0.3$).

\subsection{ARM Regulation Change}
\label{subsection: results natural expirimant}

The results in Table \ref{Table: main results regulation} take advantage of the quasi-experimental research design leveraging the regulatory change in maximum ARM ratio exposure that went into effect in January 2021. By restricting the sample to mortgages originating just before and after this change, we can examine the causal impact of higher ARM exposure while holding other factors constant. Although the ARM regulation was generally binding, the large cross-sectional differences between borrowers imply that some borrowers had an ARM ratio below the regulation limit even before the regulation change. That is, in both the treated and control samples, there are likely borrowers who were not directly impacted by the regulation. Thus, in Panel B we restrict both the control and treated groups to households whose average ARM ratio in the first quarter of 2022, right before the increase in interest rates, was in the upper two terciles of each group's distribution.\footnote{We take the upper two terciles since around 60\% of the mortgages in the period before the regulation had an approximately 33\% ARM ratio and were, therefore, likely to have been constrained by the 33\% limit.} 

\begin{table}[!htb] \centering 
  \caption{Baseline estimation - ARM regulation change} 
  \label{Table: main results regulation} 
\footnotesize
\begin{tabularx}{.9\textwidth}{@{\extracolsep{6pt}}lYY}
\\[-1.8ex]\hline 
\hline \\[-1.8ex] 
 & \multicolumn{2}{c}{Log(Deferred debit)} \\ 
\cline{2-3} 
\\[-1.8ex] 
\\[-1.8ex] & (1) & (2) \\ 
\hline \\[-1.8ex]
\multicolumn{3}{l}{\textbf{Panel A. All Households}}\\[1.4ex] 
$treat\times$Post & $-$0.013$^{**}$ & $-$0.013$^{**}$ \\ 		
  & (0.006) & (0.006) \\ 		
& & \\ 
Observations & 826,407 & 826,407 \\ 
R$^{2}$ & 0.752 & 0.752 \\ 		
Adjusted R$^{2}$ & 0.738 & 0.738 \\ 
\\[-1.8ex]
\hline \\[-1.8ex] 
\multicolumn{3}{l}{\textbf{Panel B. Top Two Terciles}}\\[1.4ex] 	
$treat\times$Post  & $-$0.018$^{***}$ & $-$0.019$^{***}$ \\ 		
  & (0.007) & (0.007) \\ 		
& & \\ 		
Observations & 550,932 & 550,932 \\ 
R$^{2}$ & 0.747 & 0.747 \\ 		
Adjusted R$^{2}$ & 0.732 & 0.732 \\ 		
& & \\ 	  
 \hline \\[-1.8ex] 
Household f.e & Y & Y \\ 		
Time f.e & Y & Y \\ 		
Credit Controls Interaction & N & Y \\ 					
\hline 
\hline \\[-1.8ex] 
\multicolumn{3}{p{0.9\textwidth}}{\textit{Notes:} This table reports the coefficient estimates of Equation (\ref{equ2}). Sample includes only borrowers who's mortgage originated between October 2020 through April 2021. Panel B excludes for both the treated and control group the lower tercile of ARM ratio. Columns represent different borrower other credit related controls. Standard errors clustered by borrower are reported in parentheses. Time period for the regression estimation is July 2021 through June 2023. $^{*}$p$<$0.1; $^{**}$p$<$0.05; $^{***}$p$<$0.01} 
\end{tabularx} 
\end{table} 

In line with the baseline results, the interaction effect between the treatment group and the post interest rate hike period indicates that mortgagors more exposed to the fluctuating ARM interest rates substantially reduced their consumption during this period. \textcolor{black}{Specifically, the results suggest that the treated group reduced consumption by 1.3\% more relative to the control. Panel B presents the results for the high ARM borrowers. As expected, the results are even stronger for those borrowers who were likely more impacted by the change in the regulation. In particular, the coefficient on the interaction effect implies the treated group reduced consumption by 1.8 \% points more relative to the control group.}

By utilizing a quasi-experimental approach, we increase confidence that the observed effect represents the causal impact of ARM exposure on consumption. Importantly, the fact that both methodologies yield similar qualitative effects further validates the significance of the channel through which adjustable-rate mortgages negatively influence spending during periods of rising interest rates.\footnote{We do not calculate an MPC for the regulatory change specification because the subsample used is quite narrow, focusing only on mortgages originated within a specific timeframe around the regulation change. This limited sample may not be representative of the broader population, making it difficult to generalize results. The primary purpose of this analysis was to confirm the direction and approximate magnitude of the effect, rather than to derive a precise MPC estimate that could be applied more broadly.} 

\begin{figure}[!htb]
\centering 
\caption{Prime rate regulation change by month}    
\includegraphics[scale=0.55]{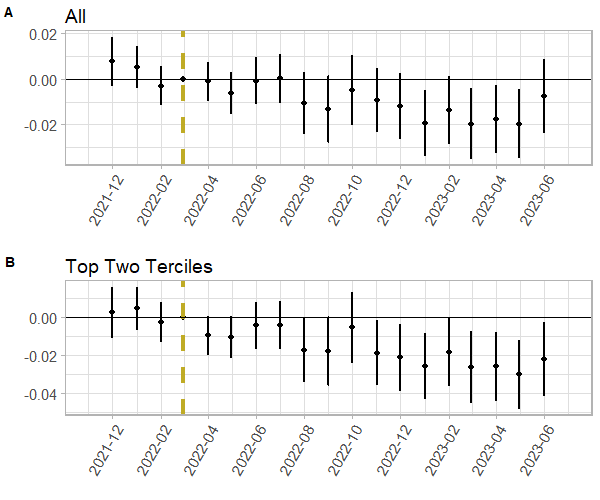}
\label{Fig dynamic regulation}
\floatfoot*{\textit{Notes}: This figure presents the dynamic estimation of Equation~\eqref{equ2}, capturing the interaction between the $\text{treat}$ variable and monthly dummy variables from December 2021 to June 2023. The estimates are presented with 90\% confidence intervals. The analysis is based on a sample of borrowers with mortgages initiated from October 2020 through April 2021. The reference point for these coefficients is set in March 2022 with its coefficient normalized to zero.} 
\end{figure}

Figure~\ref{Fig dynamic regulation} presents the dynamic specification for the DiD methodology. 
These estimates are based on the interaction between $\text{treat}$ and monthly dummies within the framework of Equation~\eqref{equ2}, accompanied by 90\% confidence intervals. The baseline for these estimates is set at the March 2022 coefficient, which is normalized to zero. As before, the dynamic analysis indicates a noticeable decline in consumption for the treated group relative to the control after March 2022, again particularly evident from August 2022 forward. This trend suggests a pronounced and enduring impact of the new, elevated interest rate environment on consumer spending behaviors, underscoring again that spending is more adversely affected as interest rates rise.

\subsection{Borrower Heterogeneity}
\label{Section: Heterogeneity}

\begin{figure}[!htb]
\centering
\includegraphics[width=1\textwidth]{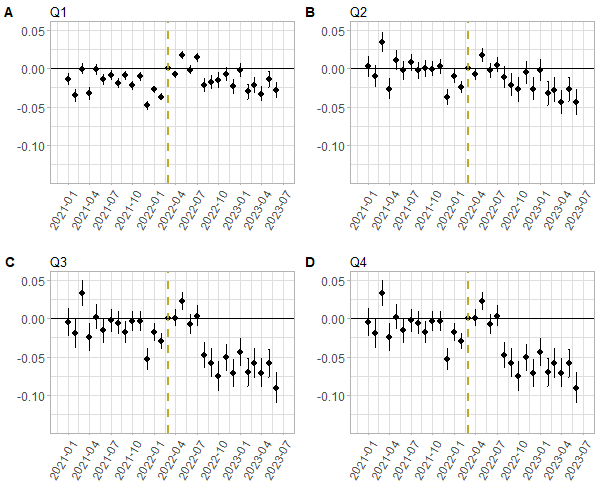}
\caption{Dynamic Impact of ARM Ratio on Debit Card Spending by Households' Payment to Spending}
\label{Fig dynamic pti}
\floatfoot*{\textit{Notes}: This figure illustrates the dynamic relationship between the ARM ratio for mortgage borrowers and the logarithm of monthly deferred debit card spending across household's average payment to differed debit card spending from January 2021 through March 2023 (\textit{pre period}). The coefficient estimates are derived from the estimation of Equation~\eqref{equ1}, incorporating interactions between the ARM ratio and monthly indicators for each month from January 2021 to June 2023. These are displayed alongside 90\% confidence intervals. Notably, the coefficient for March 2022—the month when the Bank of Israel commenced its rate hikes—is set as the baseline and normalized to zero.}
\end{figure}

In this section we first explore heterogeneity across borrowers by examining how the consumption response to rising interest rates varies depending on the household's average mortgage payment-to-spending ratio during the pre-period (January 2021 through March 2023). Specifically, we split our sample into four quartiles based on this ratio and re-estimate Equation~\eqref{equ1} separately for each group.  Table~\ref{Table: heterogeneity pti} presents the results for this heterogeneity analysis while Figure~\ref{Fig dynamic pti} presents the dynamic analysis.

The results reveal a clear pattern: households with higher payment-to-spending ratios exhibit a stronger negative consumption response to increasing ARM exposure. The interaction term coefficients become increasingly negative and statistically significant as we move up the quartiles, ranging from insignificant for the lowest quartile (Q1) to -0.091 to -0.093 for the highest quartile (Q4), all significant at the 1\% level.

The pattern of results in Table~\ref{Table: heterogeneity pti} and  Figure~\ref{Fig dynamic pti} further highlights the importance of considering household heterogeneity when assessing the impact of monetary policy on consumption through the mortgage cash-flow channel. Our findings suggest that the consumption of households with higher payment-to-income ratios is more sensitive to changes in ARM rates, as these households have less financial slack to absorb the increased debt servicing costs. This heterogeneity in consumption responses has important implications for the distributional effects of monetary policy and the potential asymmetric impacts of interest rate hikes on different segments of the population.

\begin{table}[!htb] \centering 
  \caption{Split by payment to consumption ratio} 
  \label{Table: heterogeneity pti} 
\scriptsize
\begin{tabularx}{.95\textwidth}{@{\extracolsep{6pt}}lYYYYYYYY}
\\[-2ex]\hline 
\hline \\[-2ex] 
 & \multicolumn{8}{c}{Log(Deferred debit)} \\ 
\cline{2-9} 
\\[-1.8ex] 
 & \multicolumn{2}{c}{Q1} & \multicolumn{2}{c}{Q2} & \multicolumn{2}{c}{Q3} & \multicolumn{2}{c}{Q4} \\ 		
\\[-1.8ex] & (1) & (2) & (3) & (4) & (5) & (6) & (7) & (8)\\ 				
\hline \\[-1.8ex] 
ARM ratio$\times$Post & 0.007$^{*}$ & 0.003 & $-$0.012$^{**}$ & $-$0.020$^{**}$ & $-$0.032$^{***}$ & $-$0.036$^{***}$ & $-$0.091$^{***}$ & $-$0.093$^{***}$ \\ & (0.004) & (0.004) & (0.006) & (0.006) & (0.007) & (0.007) & (0.010)  & (0.010) \\ 						
& & & & & & & &\\   
 \hline \\[-1.8ex] 
Household f.e & Y & Y & Y & Y & Y & Y & Y & Y\\ 
Time f.e  & Y & Y & Y & Y & Y & Y & Y & Y\\ 
Credit Controls Interaction & N & Y & N & Y & N & Y & N & Y \\ 
Observations & \multicolumn{2}{c}{5,259,999} & \multicolumn{2}{c}{5,259,948} & \multicolumn{2}{c}{5,259,989} & \multicolumn{2}{c}{5,259,957} \\ 
R$^{2}$ & 0.709 & 0.709 & 0.641 & 0.641 & 0.613 & 0.613 & 0.685 & 0.685 \\ 				
Adjusted R$^{2}$ & 0.698 & 0.698 & 0.628 & 0.628 & 0.599 & 0.599 & 0.672  & 0.672 \\		
\hline 
\hline \\[-1.8ex] 
\multicolumn{9}{p{0.95\textwidth}}{\textit{Notes:} This table reports the coefficient estimates of Equation (\ref{equ1}) splitting the sample by the household's average payment to differed debit card spending from January 2021 through March 2023 (\textit{pre period}). Standard errors clustered by household are reported in parentheses. Time period for the regression estimation is July 2021 through June 2023. $^{*}$p$<$0.1; $^{**}$p$<$0.05; $^{***}$p$<$0.01} 
\end{tabularx} 
\end{table}

We next investigate household income as a potential source of heterogeneity in the consumption response of mortgage borrowers to rising interest rates. To do this, we leverage the detailed household-level data available in our dataset, which includes information on the locality where each mortgage was originated. This allows us to classify households according to the socio-economic status of the locality in which they reside. Specifically, we split our sample by the socio-economic index of the household's town and re-estimate Equation~\eqref{equ1} separately for each group.

The Central Bureau of Statistics in Israel categorizes the 255 local authorities into deciles of socio-economic strength, ranging from 1 (lowest) to 10 (highest). This socio-economic index is constructed by weighing 14 different parameters, including average income, education levels, employment rates, and other relevant factors. For our analysis, we split the sample into four subgroups: households from municipalities ranked 9-10 are defined as 'High,' 7-8 as 'Medium-High,' 5-6 as 'Medium-Low,' and 3-4 as 'Low.' We exclude municipalities ranked 1-2 from the heterogeneity estimations, as there is only a small share of very low-income households with mortgages in these localities.\footnote{While mortgages are reported from the lowest-ranking municipalities, there is a greater likelihood that the municipality ranking does not accurately represent the actual income of mortgage-holding households in these areas. Jerusalem, for instance, the largest city in Israel, is ranked two on the socio-economic index. This city is highly diverse, suggesting that wealthier households likely hold a disproportionately larger share of mortgages compared to their prevalence in the overall city population. For this reason, we exclude Jerusalem from our analysis. Generally speaking, heterogeneity within a municipality is a concern across all socio-economic rankings, but it is especially pronounced in those with the lowest rankings.}

Table \ref{Table: main results heterogeneity} presents the results for these different subgroups based on the socio-economic index. The findings reveal a clear relation between the groups' socio-economic index and the magnitude of the coefficient between the ARM ratio and consumption, with the coefficient for the low- to medium-income municipalities being more than two times larger than the high-income municipalities. These results suggest that the consumption patterns of high-income households are less sensitive to changes in disposable income induced by higher mortgage payments, likely because these households can smooth consumption by drawing upon their savings.

Our findings are consistent with the economic literature on consumption smoothing and heterogeneous responses to income shocks. For example, \textcite{jappelli2010consumption} show that households with higher levels of liquid wealth are better able to insulate their consumption from income fluctuations. Similarly, \textcite{kaplan2014wealthy} demonstrate that households with substantial illiquid assets but low liquid wealth exhibit high marginal propensities to consume, making their spending more sensitive to income changes.

\begin{table}[!htb] \centering 
  \caption{Heterogeneous impact by households' municipality socio-economic index} 
  \label{Table: main results heterogeneity} 
\scriptsize
\begin{tabularx}{.95\textwidth}{@{\extracolsep{6pt}}lYYYYYYYY}
\\[-2ex]\hline 
\hline \\[-2ex] 
 & \multicolumn{8}{c}{Log(Deferred debit)} \\ 
\cline{2-9} 
\\[-1.8ex] 
 & \multicolumn{2}{c}{Low} & \multicolumn{2}{c}{Medium-Low} & \multicolumn{2}{c}{Medium-High} & \multicolumn{2}{c}{High} \\ 		
\\[-1.8ex] & (1) & (2) & (3) & (4) & (5) & (6) & (7) & (8)\\ 				
\hline \\[-1.8ex] 
ARM ratio$\times$Post & $-$0.039$^{***}$ & $-$0.051$^{***}$ & $-$0.043$^{***}$ & $-$0.049$^{***}$ & $-$0.032$^{***}$ & $-$0.039$^{***}$ & $-$0.015 & $-$0.021$^{**}$\\
& (0.014) & (0.014) & (0.006) & (0.006) & (0.004) & (0.004) &(0.009) &(0.009)\\  		
& & & & & & & &\\   
 \hline \\[-1.8ex] 
Household f.e & Y & Y & Y & Y & Y & Y & Y & Y\\ 
Time f.e  & Y & Y & Y & Y & Y & Y & Y & Y\\ 
Credit Controls Interaction & N & Y & N & Y & N & Y & N & Y \\ 
Observations & \multicolumn{2}{c}{1,378,847} & \multicolumn{2}{c}{5,969,164} & \multicolumn{2}{c}{9,578,697} & \multicolumn{2}{c}{1,737420}\\ 
R$^{2}$ & 0.743 & 0.744 & 0.753 & 0.753 & 0.753 & 0.753 & 0.729 & 0.729\\
Adjusted R$^{2}$ & 0.733 & 0.733 & 0.744 & 0.744 & 0.744 & 0.744 &0.719 &0.719 \\  							
\hline 
\hline \\[-1.8ex] 
\multicolumn{9}{p{0.95\textwidth}}{\textit{Notes:} This table reports the coefficient estimates of Equation (\ref{equ1}) splitting the sample by the socio-economic index of the households' municipality. Standard errors clustered by borrower are reported in parentheses. Time period for the regression estimation is July 2021 through June 2023. $^{*}$p$<$0.1; $^{**}$p$<$0.05; $^{***}$p$<$0.01} 
\end{tabularx} 
\end{table} 

\begin{figure}[!htb]
\centering
\includegraphics[width=1\textwidth]{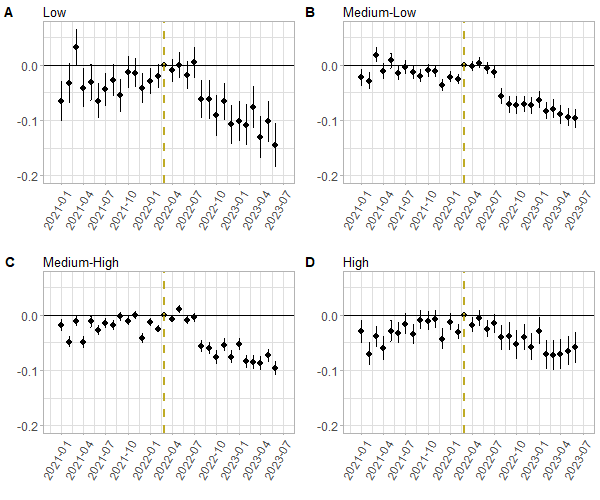}
\caption{Dynamic Impact of ARM Ratio on Debit Card Spending by Households' Socio-Economic Index}
\label{Fig dynamic heterogeneity}
\floatfoot*{\textit{Notes}: This figure illustrates the dynamic relationship between the ARM ratio for mortgage borrowers and the logarithm of monthly deferred debit card spending across socio-economic groups. The coefficient estimates are derived from the estimation of Equation~\eqref{equ1}, incorporating interactions between the ARM ratio and monthly indicators for each month from January 2021 to June 2023. These are displayed alongside 90\% confidence intervals. Notably, the coefficient for March 2022 is set as the baseline and normalized to zero.}
\end{figure}

These results are further confirmed by Figure \ref{Fig dynamic heterogeneity}, which shows the dynamic estimation for each group. Notably, the drop in consumption by June 2023 for households in Low, Medium-Low, and Medium-High municipalities is much stronger relative to households that reside in high-income towns. These findings further support the heterogeneous impact of rising interest rates on consumption, with lower-to-medium income households demonstrating greater sensitivity to changes in their mortgage payments compared to their higher-income counterparts.

\section{Robustness}
\label{Section: robustness}

\subsection{All Mortgages}
\label{subsection: robustness all}

\noindent {\textit{Selection and survival bias}}\quad Our baseline specification faces several identification concerns. First, our sample may include new mortgage borrowers who react differently to the changing interest rate environment. Specifically, borrowers who obtained mortgages in the post-period might fundamentally differ from those in the pre-period. This difference could affect the sample's structure in the post versus the pre-period, potentially leading to selection bias.\footnote{Consider, for example, in a high interest rate environment, a specific group of borrowers might be over or under-represented in new mortgages. For instance, the share of new homeowners might increase relative to switchers (people upgrading their homes). If these two groups have distinct consumption patterns, it could influence the results.}

An additional concern is that our results could be affected by people reacting to the interest rate increase by altering their mortgage structure through refinancing or paying down a portion of the mortgage.\footnote{See Bank of Israel Annual Report 2023, Chapter 4.} Moreover, it is common for a mortgage in Israel to be sanctioned with an initial lump sum followed by a series of payment facilities that are 'drawn' over a period, usually not exceeding two years, based on the house's payment schedule. Therefore, even after restricting our sample to mortgages initiated before January 2021, our sample may still include incomplete mortgages and/or changes in monthly mortgage payments due to either planned increases in the mortgage balance or borrowers' strategic reactions to the rising interest rates.

To tackle these issue, we first narrow our sample to include only those mortgage borrowers whose mortgages originated before January 2021, i.e., before the sample period (and before the ARM regulation change). Second, to address the survival bias, we further restrict our sample to borrowers whose mortgage structure and principal amount did not change by more or less than 5\% during the sample period.

Table \ref{Table: robustness all sample restriction1} displays the results for these restricted samples. In Panel A the sample is restricted to households with mortgages that started before January 202. Our findings reinforce the robustness of our results even after addressing potential selection bias by restricting the sample to mortgages originating before the sample period. Notably, the interaction term $ARMRatio\times Post$ shows a significant negative coefficient of similar magnitude, consistent with a clear shift in spending behavior after the interest rate hikes. As before, this negative effect also persists when controlling for other forms of credit.
In Panel B we restrict to mortgages that did not change the mortgage structure during the full sample period displays the results for this more narrowly defined sample. Overall, the table offers stronger evidence for the stability of our results against survival bias. Comparing the coefficient estimates from Table~\ref{Table: robustness all sample restriction1} with those in earlier tables, the interaction term "ARM ratio $\times$  Post" consistently shows a significant negative coefficient. This confirms that the spending adjustment downward is distinct among borrowers whose mortgage principals did not change after the interest rate hikes, albeit slightly less than previously observed.

\begin{table}[!htb] \centering 
  \caption{Robustness - Selection and survival bias} 
  \label{Table: robustness all sample restriction1} 
\footnotesize
\begin{tabularx}{.9\textwidth}{@{\extracolsep{6pt}}lYY}
\\[-1.8ex]\hline 
\hline \\[-1.8ex] 
 & \multicolumn{2}{c}{Log(Deferred debit)} \\ 
\cline{2-3} 
\\[-1.8ex] 
\\[-1.8ex] & (1) & (2) \\ 
\hline \\[-1.8ex]
\multicolumn{3}{l}{\textbf{Panel A. Mortgages that originated before January 2021}}\\[1.4ex] 
ARM Ratio$\times$Post & $-$0.033$^{***}$ & $-$0.042$^{***}$ \\ 				
& (0.003) & (0.003) \\ 				
& & \\ 				
Observations & 19,495,243 & 19,495,243 \\ 
R$^{2}$ & 0.764 & 0.764 \\ 				
Adjusted R$^{2}$ & 0.756 & 0.756 \\ 				
\\[-1.8ex]
\hline \\[-1.8ex] 
\multicolumn{3}{l}{\textbf{Panel B. No change in mortgage structure}}\\[1.4ex] 	
 ARM Ratio$\times$Post & $-$0.032$^{***}$ & $-$0.040$^{***}$ \\
 & (0.004) & (0.004) \\
& & \\
Observations & 12,836,930 & 12,836,930 \\ 
R$^{2}$ & 0.772 & 0.772 \\ 				
Adjusted R$^{2}$ & 0.763 & 0.763 \\ 				 
 \hline \\[-1.8ex] 
Household f.e & Y & Y \\ 		
Time f.e & Y & Y \\ 		
Credit Controls Interaction & N & Y \\ 					
\hline 
\hline \\[-1.8ex] 
\multicolumn{3}{p{0.9\textwidth}}{\textit{Notes:} This table reports the coefficient estimates of Equation (\ref{equ1}). The estimation includes households who's mortgage originated before January 2021 (Panel A) and did not change the structure of the mortgage during the estimation period (Panel B). Columns represent different borrower other credit related controls. Standard errors clustered by borrower are reported in parentheses. Time period is January 2021 through June 2023. $^{*}$p$<$0.1; $^{**}$p$<$0.05; $^{***}$p$<$0.01} 
\end{tabularx} 
\end{table}

\vspace{0.2cm} \noindent {\textit{Empirical specification}}\quad 
We test the robustness of our results to a number of alternative variable specification.

First, using the $Post$ dummy in our baseline estimation effectively implies that we are estimating the average response of mortgage borrowers' consumption during the period of rising interest rates. We believe this specification is appropriate as it captures both the accumulating increase in the policy rate as well as the expectations and overall environment during this period. Alternatively, one can use the actual monthly policy rate, namely $MP$, instead of $Post$. Second, we test the robustness of the results for using the log of the actual monthly mortgage payments of each households instead of the ARM ratio. The focus now is on estimating the impact of higher mortgage payments, in percent, on the consumption of borrowers, again in terms of deferred debit card spending. 
Finally, we use a dummy variable equal to one for "High ARM" households defined as households whose average pre period ARM ratio was above the sample median instead of the continuous ARM ratio. 

Table \ref{Table: robustness Empirical specification} presents the results for these alternative specifications. Panel A presents the results using the monetary policy rate. The interpretation of the coefficients is that a one percentage point increase in the BOI policy rate is associated with a reduction between 1.7 to 1.9 basis points for every one percentage point higher ARM ratio. Note that while the total policy rate increase during the period was 4.65\% the table estimates an average response. Since the policy rate averaged 2.68\% during the post period the results imply an average reduction of between 4.5 to 5 basis points reduction during the post period, overall consistent with the baseline results.

Panel B report the results of the estimation using the contemporaneous log of the monthly mortgage payments instead of the ARM ratio. Note that in this specification the coefficient on the $\log(Mortgage \: Payment)$ without the interaction does not drop since the monthly mortgage payments changes over time and is thus not absorbed by the household fixed effect. The coefficients of the interaction implies that in the post period a 1\% increase in mortgage payments leads to 1.8-1.9 basis points decrease in deferred debit spending. Recall that payments increase on average by around 15\%. Thus the results imply an overall reduction in consumption in the post period for the average household of 0.29\% which is overall consistent with overall average reduction of 0.4\% estimated in the baseline specification. 

Panel C reports the results for "High ARM" dummy. The results implies that the high ARM mortgage households reduced consumption by 0.2-0.25 \% more relative to low ARM households which is somewhat weaker but overall consistant with the baseline estimation.

 \begin{table}[!htb] \centering 
  \caption{Robustness - Empirical specification} 
  \label{Table: robustness Empirical specification} 
\footnotesize
\begin{tabularx}{.9\textwidth}{@{\extracolsep{6pt}}lYY}
\\[-1.8ex]\hline 
\hline \\[-1.8ex] 
 & \multicolumn{2}{c}{Log(Deferred debit)} \\ 
\cline{2-3} 
\\[-1.8ex] 
\\[-1.8ex] & (1) & (2) \\ 
\hline \\[-1.8ex]
\multicolumn{3}{l}{\textbf{Panel A.  Monetary policy rate}}\\[1.4ex] 
 ARM Ratio$\times$MP & $-$0.017$^{***}$ & $-$0.019$^{***}$ \\
 & (0.001) & (0.001) \\
& & \\ 							
R$^{2}$ & 0.762 & 0.762 \\ 				
Adjusted R$^{2}$ & 0.753 & 0.753 \\ 			
\\[-1.8ex]
\hline \\[-1.8ex] 
\multicolumn{3}{l}{\textbf{Panel B. Mortgage payment}}\\[1.4ex] 
$\log(Mortgage \: Payment)$ & 0.062$^{***}$ & 0.062$^{***}$ \\
& (0.001) & (0.001) \\
$\log(Mortgage \: Payment)\times$Post & $-$0.019$^{***}$ & $-$0.018$^{***}$ \\
  & (0.001) & (0.001) \\
& & \\ 
R$^{2}$ & 0.762 & 0.762 \\ 				
Adjusted R$^{2}$ & 0.753 & 0.753 \\ 			
\\[-1.8ex]
\hline \\[-1.8ex] 				
\multicolumn{3}{l}{\textbf{Panel C. Dummy for high ARM ratio}}\\[1.4ex] 	
High ARM$\times$Post & $-$0.020$^{***}$ & $-$0.025$^{***}$ \\ 				
  & (0.001) & (0.001) \\ 				
& & \\ 				
R$^{2}$ & 0.762 & 0.762 \\ 				
Adjusted R$^{2}$ & 0.753 & 0.753 \\ 	
\\[-1.8ex]				
\hline \\[-1.8ex] 
Household f.e & Y & Y \\ 				
Time f.e & Y & Y \\ 				
Credit Controls Interaction & N & Y \\ 
Observations & 21,287,084 & 21,287,084 \\ 	
\hline 
\hline \\[-1.8ex] 
\multicolumn{3}{p{0.9\textwidth}}{\textit{Notes:} This table reports the coefficient estimates of Equation (\ref{equ1}) using alternative specifications. Panel A uses the monthly level of the BOI policy rate (MP) instead of the $post$ indicator. Panel B uses the log of the contemporaneous monthly mortgage payment instead of the ARM ratio. In Panel C ARM ratio is replaced by a dummy variable equal to one if the household's ARM ratio is above the full sample median. Columns represent different borrower other credit related controls. Standard errors clustered by borrower are reported in parentheses. Time period is January 2021 through June 2023. $^{*}$p$<$0.1; $^{**}$p$<$0.05; $^{***}$p$<$0.01} 
\end{tabularx} 
\end{table}

\vspace{0.2cm} \noindent {\textit{Mortgage structure}}\quad
In our baseline specification, we focus solely on the ARM ratio as the key explanatory variable, without controlling for other mortgage tracks such as the share of fixed-CPI linked or medium-fixed rate mortgages. Including these additional controls is not straightforward, as they may be considered "bad controls" in the sense that they are simultaneously determined with the ARM ratio. The choice of mortgage tracks is typically made jointly by borrowers, implying that the ratios of different tracks within a mortgage are inherently correlated. Consequently, directly controlling for the share of other tracks could potentially introduce endogeneity issues and bias our estimates of the ARM ratio's impact on consumption.

Table~\ref{Table: robustness Mortgage structure} presents a robustness check addressing potential concerns about the impact of mortgage structure on our findings. To mitigate these issues, we restricted our sample to mortgages originated between July 2018 and December 2020, with a CPI-linked component less than 5\% of the total mortgage. This approach reduces sensitivity to CPI-linked components and minimizes potential impacts from medium-term interest rate changes in other tracks, albeit at the cost of a reduced sample size. The results strongly confirm the robustness of our findings. The effect of ARM ratio on deferred debt remains statistically significant and increases substantially in magnitude to -0.103. This amplified effect, significant at the 1\% level, can be attributed to the relatively high exposure to ARM within this subsample.

The explanation for this stronger effect lies in the increased stability of cashflow of the non-ARM part of the mortgages in this restricted sample. In our baseline regression, the cashflow of some part of the non-ARM mortgage is also influenced by changes in the interest rates, leading to a slight decrease in consumption that attenuates the overall effect. Consequently, it's logical that the coefficient would be larger in this robustness check. Our conclusion is that indexation and medium-fix components do have an effect, which weakens the impact in the basic regression. This suggests that our initial results may, if anything, underestimate the true effect of ARM on deferred debt. These findings not only reinforce our main conclusions about the impact of ARM on deferred debt but also effectively address concerns about confounding effects of mortgage structure.

\begin{table}[!htb] \centering 
  \caption{Robustness - Mortgage structure} 
  \label{Table: robustness Mortgage structure} 
\footnotesize
\begin{tabularx}{.9\textwidth}{@{\extracolsep{6pt}}lYY}
\\[-1.8ex]\hline 
\hline \\[-1.8ex] 
 & \multicolumn{2}{c}{Log(Deferred debit)} \\ 
\cline{2-3} 
\\[-1.8ex] 
\\[-1.8ex] & (1) & (2) \\ 
\hline \\[-1.8ex]
ARM Ratio$\times$Post & $-$0.091$^{***}$ & $-$0.103$^{***}$ \\ 
& (0.016) & (0.016) \\				
\\[-1.8ex]							
\hline \\[-1.8ex] 
Household f.e & Y & Y \\ 				
Time f.e & Y & Y \\ 				
Credit Controls Interaction & N & Y \\ 
Observations & 982,396 & 982,396 \\ 	
R$^{2}$ & 0.740 & 0.740 \\ 				
Adjusted R$^{2}$ & 0.730 & 0.731 \\ 	
\hline 
\hline \\[-1.8ex] 
\multicolumn{3}{p{0.9\textwidth}}{\textit{Notes:} This table reports the coefficient estimates of Equation (\ref{equ1}) restricting the sample to mortgages taken between July 2018 through December 2020 with less then 5\% of the mortgage linked to the CPI index. Columns represent different borrower other credit related controls. Standard errors clustered by borrower are reported in parentheses. Time period is January 2021 through June 2023. $^{*}$p$<$0.1; $^{**}$p$<$0.05; $^{***}$p$<$0.01} 
\end{tabularx} 
\end{table} 

\vspace{0.2cm} \noindent {\textit{Using Monetary Policy Shocks}}\quad
Table~\ref{Table: robustness Empirical specification2}  presents further robustness checks that validate our main findings regarding the mortgage cash-flow channel. The analysis in this table replaces the binary post-period indicator used in the baseline specification with measures of monetary policy shocks 
providing a more granular examination of how consumption responds to interest rate changes.

In Panel A, we interact the ARM ratio with the cumulative monetary policy shocks over the entire sample period. The negative and statistically significant coefficient estimates of -0.075 and -0.085 confirm that households with higher exposure to adjustable-rate mortgages experienced a more pronounced reduction in consumption in response to the accumulated interest rate hikes. Specifically, a one percentage point increase in the cumulative monetary policy shock leads to an additional 7.5 to 8.5 basis point decline in deferred debit spending for every one percentage point higher ARM ratio. This finding is consistent with our central hypothesis that the cash-flow channel amplifies the transmission of monetary policy to household spending.

Panel B of Table~\ref{Table: robustness Empirical specification2} takes a slightly different approach, using the actual month-to-month monetary policy shocks and specifying the dependent variable as the log change in deferred debit card spending. This specification captures the immediate consumption response to each incremental interest rate change. The negative and significant interaction term coefficients of -0.072 and -0.069 align with our key result, demonstrating that the mortgage cash-flow channel intensifies the impact of rate increases on consumption, particularly for households with high ARM ratios. The interpretation is that a one percentage point monetary policy shock results in an additional 6.9 to 7.2 basis point decrease in the monthly change in deferred debit spending for every one percentage point higher ARM ratio.

 \begin{table}[!htb] \centering 
  \caption{Robustness - Empirical Specification II} 
  \label{Table: robustness Empirical specification2} 
\footnotesize
\begin{tabularx}{.9\textwidth}{@{\extracolsep{6pt}}lYY}
\\[-1.8ex]\hline 
\hline \\[-1.8ex] 
 & \multicolumn{2}{c}{Log(Deferred debit)} \\ 
\cline{2-3} 
\\[-1.8ex] 
\\[-1.8ex] & (1) & (2) \\ 
\hline \\[-1.8ex]
\multicolumn{3}{l}{\textbf{Panel A.  Monetary policy Shocks (Cumulative)}}\\[1.4ex] 
 ARM Ratio$\times$MP Shock & $-$0.075$^{***}$ & $-$0.085$^{***}$ \\
 & (0.005) & (0.005) \\
& & \\ 							
R$^{2}$ & 0.762 & 0.762 \\ 				
Adjusted R$^{2}$ & 0.753 & 0.753 \\ 
Observations & 21,287,084 & 21,287,084 \\ 			
\\[-1.8ex]
\hline \\[-1.8ex] 
 & \multicolumn{2}{c}{Log Change (Deferred debit)} \\ 
\cline{2-3} 
\\[-1.8ex] 
\\[-1.8ex] & (1) & (2) \\ 
\hline \\[-1.8ex]
\multicolumn{3}{l}{\textbf{Panel B. Monetary Policy Shocks}}\\[1.4ex] 
 ARM Ratio$\times$MP Shock & $-$0.072$^{***}$ & $-$0.069$^{***}$ \\
 & (0.007) & (0.007) \\
& & \\ 							
R$^{2}$ & 0.032 & 0.032 \\ 				
Adjusted R$^{2}$ & -0.007 & -0.007 \\ 
Observations & 20,501,522 & 20,501,522 \\ 	
\\[-1.8ex]				
\hline \\[-1.8ex] 
Household f.e & Y & Y \\ 				
Time f.e & Y & Y \\ 				
Credit Controls Interaction & N & Y \\ 	
\hline 
\hline \\[-1.8ex] 
\multicolumn{3}{p{0.9\textwidth}}{\textit{Notes:} This table reports the coefficient estimates of Equation (\ref{equ1}) using monetary policy (MP) shocks instead $post$ indicator. Panel A uses the cumulative MP shocks through the full sample . Panel B uses the MP shocks with a log change specification. Columns represent different borrower other credit related controls. Standard errors clustered by borrower are reported in parentheses. Time period is January 2021 through June 2023. $^{*}$p$<$0.1; $^{**}$p$<$0.05; $^{***}$p$<$0.01} 
\end{tabularx} 
\end{table}

Table~\ref{Table: robustness all alternative controls} presents robustness checks for our main findings using alternative control variables. While our baseline specification uses dummy variables to control for the presence of other types of debt, such as consumer loans, overdrafts, and credit card debt, this table employs continuous variables representing the actual balances of these debt types (in thousands of NIS) to capture the intensive margin effect.

The interaction term between the ARM ratio and the post-period dummy remains negative and statistically significant, with coefficient estimates ranging from -0.0328 to -0.036. This finding confirms that our main result – households with higher exposure to adjustable-rate mortgages reduced their consumption more during the period of rising interest rates – is robust to the choice of control variables. The magnitude of the coefficients suggests that a one percentage point higher ARM ratio is associated with an additional 3.28 to 3.6 basis point decline in deferred debit spending during the post-period.

The coefficients on the interaction terms between the post-period dummy and the continuous debt variables are also negative and statistically significant, albeit with smaller magnitudes compared to the ARM ratio interaction. A one thousand NIS increase in consumer loans, overdrafts, or credit card balances is associated with an additional 0.01 to 0.03 basis point decrease in deferred debit spending during the post-period. 

\begin{table}[!htb] \centering 
  \caption{Robustness - alternative controls} 
  \label{Table: robustness all alternative controls} 
\footnotesize
\begin{tabularx}{.9\textwidth}{@{\extracolsep{6pt}}lYY}
\\[-1.8ex]\hline 
\hline \\[-1.8ex] 
 & \multicolumn{2}{c}{Log(Deferred debit)} \\ 
\cline{2-3} 
\\[-1.8ex] 
\\[-1.8ex] & (1) & (2) \\ 
\hline \\[-1.8ex]
ARM ratio$\times$Post & $-$0.036$^{***}$ & $-$0.0328$^{***}$\\ 
  & (0.003) & (0.003) \\ 
  
  Consumer loans$\times$Post &  & $-$0.0001$^{***}$ \\ 
  &  &  (0.00001) \\ 
  Overdraft$\times$Post &   & $-$0.0003$^{***}$ \\ 
  &  & (0.00001) \\ 
  Credit card$\times$Post &  & $-$0.0002$^{***}$ \\ 
  &  &  (0.00001) \\ 
 \hline \\[-1.8ex] 
Household f.e & Y & Y \\ 
Time f.e  & Y & Y \\ 
Credit Controls Interaction & N & Y \\ 
Observations & 21,287,084 & 21,287,084\\ 			
R$^{2}$ & 0.762 & 0.762  \\ 
Adjusted R$^{2}$ & 0.753 & 0.753 \\ 
\hline 
\hline \\[-1.8ex] 
\multicolumn{3}{p{0.9\textwidth}}{\textit{Notes:} This table reports the coefficient estimates of Equation (\ref{equ1}), using continuous control variables (balances in thousands) instead of dummy variables. Standard errors clustered by borrower are reported in parentheses. Time period is January 2021 through June 2023. $^{*}$p$<$0.1; $^{**}$p$<$0.05; $^{***}$p$<$0.01} 
\end{tabularx} 
\end{table}

\begin{figure}[!htb]
\centering
\includegraphics[width=0.9\textwidth]{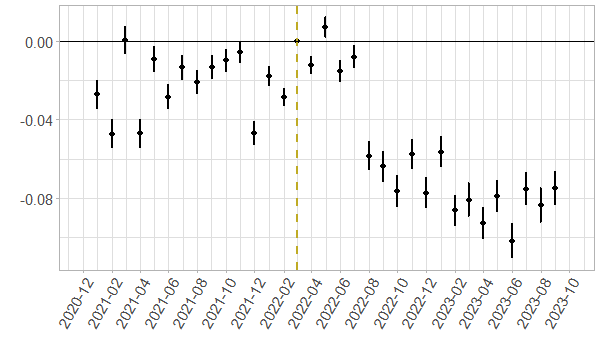}
\caption{Dynamic Impact  - Longer Sample}
\label{Fig dynamic main long}
\floatfoot*{\textit{Notes}: This figure reports the dynamic relationship between mortgage borrowers' ARM rate and the logarithm of monthly deferred debit card spending. The coefficient estimates presented come from the estimation of Equation~\eqref{equ1}, which includes interactions between "ARM ratio" and monthly dummies for each month from February 2021 to September 2023, displayed with 90\% confidence bands. The coefficient for April 2022, the month when the Bank of Israel started hiking rates, is normalized to zero.}
\end{figure}

Finally, Figure~\ref{Fig dynamic main long} extends our dynamic analysis of the relationship between mortgage borrowers' ARM exposure and their deferred debit card spending by including an additional three months of data, covering the period from February 2021 to September 2023.

The figure plots the coefficient estimates from the dynamic version of Equation~\eqref{equ1}, which interacts the ARM ratio with monthly dummy variables, along with their 90\% confidence intervals. As before, the coefficient for March 2022, the month when the BOI began hiking interest rates, is normalized to zero.

The results confirm and reinforce our main findings. The negative impact of higher ARM exposure on consumption becomes more evident from August 2022 onward, coinciding with the BOI's acceleration of interest rate increases. By September 2023, the end of our extended sample period, the interaction coefficient reaches approximately -0.08, implying that a one percentage point higher ARM ratio is associated with an additional 8 basis point decline in deferred debit card spending compared to April 2022.

The inclusion of three additional months of data in Figure 10 allows us to track the evolution of the consumption response over a longer time horizon. The persistent and increasing magnitude of the negative interaction coefficients throughout the extended period underscore the enduring impact of the mortgage cash-flow channel on household spending. These findings provide further support for our conclusion that adjustable-rate mortgage exposure plays a crucial role in amplifying the transmission of monetary policy to consumption, particularly during periods of rising interest rates.

\subsection{Prime Regulation Change}
\label{subsection: robustness prime}

Table \ref{Table: robustness regulation} presents our robustness result for the DiD specification. In Panel A we use a restricted sample of borrowers who did not change their mortgage structure during the sample period. In panel B we drop mortgages that originated in January, the month that the regulation change happened. In Panels C and D we do the same restrictions for the upper ARM terciles sample.

As shown in the table, and similar to the previous section, restricting the sample to borrowers who did not change their mortgages does not alter our conclusions. The coefficient on the interaction term remains in the vicinity of 3 basis points, and its magnitude is slightly lower once we include our set of control variables.

 \begin{table}[!htb] \centering 
  \caption{Robustness - ARM regulation change} 
  \label{Table: robustness regulation} 
\scriptsize
\begin{tabularx}{.9\textwidth}{@{\extracolsep{6pt}}lYY}
\\[-1.8ex]\hline 
\hline \\[-1.8ex] 
 & \multicolumn{2}{c}{Log(Deferred debit)} \\ 
\cline{2-3} 
\\[-1.8ex] 
\\[-1.8ex] & (1) & (2) \\ 
\hline \\[-1.8ex]
\multicolumn{3}{l}{\textbf{Panel A. Survival bias}}\\[1.4ex] 
treat$\times$Post & $-$0.012$^{*}$ & $-$0.012$^{*}$ \\ 		
& (0.007) & (0.007) \\ 		
& & \\ 		
Observations & 611,616 & 616,616 \\ 
R$^{2}$ & 0.753 & 0.753 \\ 		
Adjusted R$^{2}$ & 0.739 & 0.739 \\ 				
\\[-1.8ex]
\hline \\[-1.8ex] 
\multicolumn{3}{l}{\textbf{Panel B. Dropping January}}\\[1.4ex] 
treat$\times$Post & $-$0.011$^{*}$ & $-$0.011$^{*}$ \\ 		
& (0.006) & (0.006) \\ 		
& & \\ 		
Observations & 748,327 & 748,327 \\ 
R$^{2}$ & 0.752 & 0.753 \\ 		
Adjusted R$^{2}$ & 0.738 & 0.738 \\ 			
\\[-1.8ex]
\hline \\[-1.8ex] 				
\multicolumn{3}{l}{\textbf{Panel C. Survival bias - Top Two Terciles}}\\[1.4ex] 	
treat$\times$Post & $-$0.019$^{**}$ & $-$0.020$^{**}$ \\ 		
& (0.008) & (0.008) \\ 		
&  &  \\ 
Observations & 434,657 & 434,657 \\ 
R$^{2}$ & 0.746 & 0.746 \\ 		
Adjusted R$^{2}$ & 0.731 & 0.731 \\ 		
\\[-1.8ex]
\hline \\[-1.8ex] 				
\multicolumn{3}{l}{\textbf{Panel D. Dropping January - Top Two Terciles}}\\[1.4ex] 	
treat$\times$Post & $-$0.018$^{**}$ & $-$0.018$^{**}$ \\ 		
& (0.007) & (0.007) \\ 		
Observations & 503,107 & 503.107 \\ 
R$^{2}$ & 0.747 & 0.747 \\ 		
Adjusted R$^{2}$ & 0.733 & 0.733 \\ 		
\\[-1.8ex]				
\hline \\[-1.8ex] 
Household f.e & Y & Y \\ 				
Time f.e & Y & Y \\ 				
Credit Controls Interaction & N & Y \\ 	
\hline 
\hline \\[-1.8ex] 
\multicolumn{3}{p{0.9\textwidth}}{\textit{Notes:} This table reports the coefficient estimates of Equation (\ref{equ2}). Sample includes only borrowers who's mortgage originated between October 2020 through April 2021. Panel A include only households who's mortgage structure did not change during the estimation period. Panel B excludes households with mortgages that were taken in January 2021. Panel C excludes for both the treated and control group the lower tercile of ARM ratio. Panel C excludes for both the treated and control group the lower tercile of ARM ratio households with mortgages that were taken in January 2021. Columns represent different borrower other credit related controls. Standard errors clustered by household are reported in parentheses. Time period is December 2021 through June 2023. $^{*}$p$<$0.1; $^{**}$p$<$0.05; $^{***}$p$<$0.01} 
\end{tabularx} 
\end{table}

To address potential concerns about spurious findings in our Difference-in-Differences (DiD) analysis, we additionally conducted a standard placebo test. This test involved running the same DiD specification but with an arbitrary cutoff date set to January of four previous years (2017-2020), during which there were no changes in regulation. Specifically, in each panel estimation is performed only for borrowers whose mortgage originated between October 2016/2017/2018/2019 through April 2017/2018/2019/2020. 

The results of our placebo test are documented in Table~\ref{Table: robustness placebo}. The results reveal largely  insignificant results. 
This lack of significance in the placebo test provides additional confidence in the reliability of our DiD methodology and its estimations, bolstering the credibility of our findings.

\begin{table}[!htb] \centering 
  \caption{Robustness - prime regulation change placebo test} 
  \label{Table: robustness placebo} 
\footnotesize
\begin{tabularx}{.9\textwidth}{@{\extracolsep{6pt}}lYY}
\\[-1.8ex]\hline 
\hline \\[-1.8ex] 
 & \multicolumn{2}{c}{Log(Deferred debit)} \\ 
\cline{2-3} 
\\[-1.8ex] 
\\[-1.8ex] & (1) & (2) \\ 
\hline \\[-1.8ex]
\multicolumn{3}{l}{\textbf{Panel A: January 2020}} \\ 
treat$\times$Post & 0.014$^{**}$ & 0.013$^{**}$ \\ 		
& (0.006) & (0.006) \\ 		
\\[-1.8ex] 	
Observations & 721,892 & 721,892 \\ 	
R$^{2}$ & 0.750 & 0.750 \\ 		
Adjusted R$^{2}$ & 0.736 & 0.736 \\ 		
\hline \\[-1.8ex]
\multicolumn{3}{l}{\textbf{Panel B: January 2019}} \\ 
treat$\times$Post & $-$0.007 & $-$0.007 \\ 		
& (0.007) & (0.007) \\ 		
\\[-1.8ex] 		
Observations & 523,619 & 523,619 \\ 
R$^{2}$ & 0.759 & 0.759 \\ 		
Adjusted R$^{2}$ & 0.745 & 0.745 \\ 		
\hline \\[-1.8ex]
\multicolumn{3}{l}{\textbf{Panel C: January 2018}} \\ 
treat$\times$Post & $-$0.008 & $-$0.008 \\ 		
& (0.007) & (0.007) \\ 		
\\[-1.8ex] 		
Observations & 439,846 & 439,846 \\ 
R$^{2}$ & 0.768 & 0.768 \\ 		
Adjusted R$^{2}$ & 0.754 & 0.754 \\ 		
\hline \\[-1.8ex]
\multicolumn{3}{l}{\textbf{Panel D: January 2017}} \\  
treat$\times$Post & $-$0.009 & $-$0.009 \\
& (0.008) & (0.008) \\
\\[-1.8ex]		
Observations & 333,003 & 333,003 \\ 
R$^{2}$ & 0.779 & 0.779 \\ 		
Adjusted R$^{2}$ & 0.766 & 0.766 \\ 		
\hline \\[-1.8ex] 
Household f.e & Y & Y \\ 				
Time f.e & Y & Y \\ 				
Credit Controls Interaction & N & Y \\ 	
\hline 
\hline \\[-1.8ex] 
\multicolumn{3}{p{0.9\textwidth}}{\textit{Notes:} This table reports the coefficient estimates of the placebo test for the ARM regulation that happened in January 2021. Estimation is exactly as described in Section \ref{subsection: prime regulation} but with different dates for determining the treatment and control. Specifically, in each panel estimation is performed only for borrowers whose mortgage originated between October 2016/2017/2018/2019 through April 2017/2018/2019/2020. Standard errors clustered by borrower are reported in parentheses. Time period is December 2021 through June 2023. $^{*}$p$<$0.1; $^{**}$p$<$0.05; $^{***}$p$<$0.01.} 
\end{tabularx} 
\end{table} 


Table~\ref{Table: robustness window} presents additional robustness checks for our DiD analysis where we test the sensitivity of our results to different time windows around the regulation change date for defining treatment and control groups. Columns 1 and 2 use a narrow one-month window, including only borrowers whose mortgages originated between December 2020 and February 2021. The coefficient on the interaction term between the treatment dummy and the post-period dummy is negative but not statistically significant, with values of -0.011 and -0.010.

Columns 3 and 4 expand the window to two months, covering borrowers whose mortgages originated between November 2020 and March 2021. The interaction term coefficients become more negative and statistically significant at the 5\% and 10\% levels, with values of -0.016 in both specifications.

Columns 5 and 6 further extend the window to four months, including borrowers whose mortgages originated between September 2020 and May 2021. The interaction term coefficients remain negative and are now statistically significant at the 1\% level, with a value of -0.018 in both specifications.

The pattern of results suggests that the impact of the ARM ratio regulation change on consumption becomes more pronounced as we widen the time window around the change date. This finding is consistent with the idea that it takes time for borrowers to adjust their mortgage choices and consumption behavior in response to the regulatory change. The statistical significance and magnitude of the coefficients in the four-month window specification (columns 5 and 6) are similar to our main DiD results, supporting the robustness of our findings.

\begin{table}[!htb] \centering 
  \caption{Robustness - ARM Regulation Change - Alternative Time Windows} 
  \label{Table: robustness window} 
\scriptsize
\begin{tabularx}{.95\textwidth}{@{\extracolsep{6pt}}lYYYYYY}
\\[-2ex]\hline 
\hline \\[-2ex] 
 & \multicolumn{6}{c}{Log(Deferred debit)} \\ 
\cline{2-7} 
\\[-1.8ex] 
 & \multicolumn{2}{c}{One Month} & \multicolumn{2}{c}{Two Months} & \multicolumn{2}{c}{Four Months}\\ 		
\\[-1.8ex] & (1) & (2) & (3) & (4) & (5) & (6)\\ 				
\hline \\[-1.8ex] 
treat$\times$Post & $-$0.011 & $-$0.010 & $-$0.016$^{**}$ & $-$0.016$^{*}$ & $-$0.018$^{***}$ & $-$0.018$^{***}$\\
& (0.009) & (0.009) & (0.007) & (0.007) & (0.005) & (0.005) \\  	
& & & & & & \\   
 \hline \\[-1.8ex] 
Household f.e & Y & Y & Y & Y & Y & Y\\ 
Time f.e  & Y & Y & Y & Y & Y & Y\\ 
Credit Controls Interaction & N & Y & N & Y & N & Y\\ 
Observations & \multicolumn{2}{c}{350,405} & \multicolumn{2}{c}{598,481} & \multicolumn{2}{c}{1,078,483} \\ 
R$^{2}$ & 0.748 & 0.749 & 0.747 & 0.747 & 0.751 & 0.751\\
Adjusted R$^{2}$ & 0.734 & 0.734 & 0.733 & 0.733 & 0.737 & 0.737\\  							
\hline 
\hline \\[-1.8ex] 
\multicolumn{7}{p{0.95\textwidth}}{\textit{Notes:} This table reports the coefficient estimates of Equation \ref{equ2}. Columns 1-2 include only borrowers who's mortgage originated between December 2020 through February 2021. Columns 3-4 include only borrowers who's mortgage originated between November 2020 through March 2021. Columns 5-6 include only borrowers who's mortgage originated between September 2020 through May 2021. Standard errors clustered by household are reported in parentheses. Time period for the regression estimation is December 2021 through June 2023. $^{*}$p$<$0.1; $^{**}$p$<$0.05; $^{***}$p$<$0.01} 
\end{tabularx} 
\end{table}

\section{Conclusions}
\label{Section: Conclusions}

This paper examines the mortgage cash-flow channel of monetary policy transmission. Using detailed household-level data and variations in adjustable-rate mortgage exposure, we show that increased debt servicing costs from interest rate increases lead to notable reductions in consumption spending. Our research indicates that households with greater exposure to rate fluctuations through adjustable-rate mortgages reduced their consumption by an additional 3.6\% following a 465 basis point increase in the monetary policy rate. We also observe that this effect is not constant - it is minimal at the start of the tightening period and becomes more pronounced towards the end of our sample. Lastly, we find that the spending response to an interest rate increase through exposure to adjustable-rate mortgages is strongest among middle to lower-income households and those with high payment-to-consumption ratios, suggesting potential distributional effects of the mortgage cash-flow channel.

These results have several key implications. First, they quantify a mechanism of monetary policy transmission that has received less empirical scrutiny compared to other channels. By establishing a causal relationship between adjustable-rate mortgage exposure and consumption responses, we deepen understanding of how interest rates influence economic activity. Second, our findings demonstrate significant heterogeneity in transmission effects, with low and middle-income groups and those with high payment-to-consumption ratios exhibiting greater sensitivity. This suggests monetary policy changes mediated through consumer debt and cash flows may exacerbate inequality. Accounting for these distributional impacts should be an important consideration for central banks, and further investigation is warranted. Finally, the results highlight how regulatory choices in mortgage markets can dramatically affect policy transmission. The shift allowing a higher share of adjustable-rate mortgages meaningfully intensified the consumption effect of rate hikes. This finding indicates that greater coordination between central banks and financial supervision agencies is warranted.

Our paper highlights the importance of considering household balance sheets when making monetary policy decisions. Changes in mortgage rates significantly affect borrowers' demand. As debt service burdens increase in many countries, taking these dynamics into account when setting policy will be important for stabilizing output and inflation, as well as maintaining financial stability.

\newpage
\singlespace
\printbibliography

\clearpage
\appendix

\end{document}